\RequirePackage{fix-cm}
\documentclass[twocolumn,epjc3]{svjour3}  
\smartqed  
\RequirePackage{graphicx}
\graphicspath{{img/}} 

\RequirePackage{xspace}
\journalname{Eur. Phys. J. C}

\RequirePackage{amsmath}
\newcommand{\mmu}{\ensuremath{\langle\,\mu\,\rangle}\xspace}
\newcommand{\pt}{\ensuremath{p_\text{T}\xspace}}

\RequirePackage{floatrow}
\newfloatcommand{capbtabbox}{table}[][\FBwidth]

\RequirePackage{booktabs}
\RequirePackage{xcolor}
\definecolor{darkgreen}{rgb}{0.0, 0.4, 0.0}
\RequirePackage{hyperref}
\hypersetup{
    linktocpage,
     colorlinks,
     citecolor=darkgreen,
     linkcolor= darkgreen,
     urlcolor=darkgreen
}

\RequirePackage{comment}
\RequirePackage{lineno}
\RequirePackage{todonotes}
\begin{document}
\title{Performance of a Geometric Deep Learning Pipeline for HL-LHC Particle Tracking
}

\author{
Xiangyang Ju\thanksref{a1,e1} \and
{Daniel} {Murnane}\thanksref{a1} \and
{Paolo} {Calafiura}\thanksref{a1} \and
{Nicholas} {Choma}\thanksref{a1} \and
{Sean} {Conlon}\thanksref{a1} \and
{Steven} {Farrell}\thanksref{a1} \and
{Yaoyuan} {Xu}\thanksref{a1} \and
{Maria} {Spiropulu}\thanksref{a2} \and
{Jean-Roch} {Vlimant}\thanksref{a2} \and
{Adam} {Aurisano}\thanksref{a3} \and
{Jeremy} {Hewes}\thanksref{a3} \and
{Giuseppe} {Cerati}\thanksref{a4} \and
{Lindsey} {Gray}\thanksref{a4} \and
{Thomas} {Klijnsma}\thanksref{a4} \and
{Jim} {Kowalkowski}\thanksref{a4} \and
{Markus} {Atkinson}\thanksref{a5} \and
{Mark} {Neubauer}\thanksref{a5} \and
{Gage} {DeZoort}\thanksref{a6} \and
{Savannah} {Thais}\thanksref{a6} \and
{Aditi} {Chauhan}\thanksref{a7} \and
{Alex} {Schuy}\thanksref{a7} \and
{Shih-Chieh} {Hsu}\thanksref{a7} \and
{Alex} {Ballow}\thanksref{a8} \and
{Alina} {Lazar}\thanksref{a8} 
}

\xspace

\thankstext{e1}{\email{xju@lbl.gov}}
\institute{
\label{a1}
Lawrence Berkeley National Laboratory, Berkeley, CA, USA
\and
\label{a2}
California Institute of Technology, Pasadena, CA USA
\and
\label{a3}
University of Cincinnati, Cincinnati, OH USA
\and
\label{a4}
Fermi National Accelerator Laboratory, Batavia, IL USA
\and
\label{a5}
University of Illinois at Urbana-Champaign, Urbana, IL, USA
\and
\label{a6}
Princeton University, Princeton, NJ, USA
\and
\label{a7}
University of Washington, Seattle, WA, USA
\and 
\label{a8}
Youngstown State University, Youngstown, OH, USA
}

\maketitle

\abstract{
The Exa.TrkX project has applied geometric learning concepts such as metric learning and graph neural networks to HEP particle tracking. Exa.TrkX's tracking pipeline groups detector measurements to form track candidates and filters them. The pipeline, originally developed using the TrackML dataset (a simulation of an LHC-inspired tracking detector), has been demonstrated on other detectors, including DUNE Liquid Argon TPC and CMS High-Granularity Calorimeter. This paper documents new developments needed to study the physics and computing performance of the Exa.TrkX pipeline on the full TrackML dataset, a first step towards validating the pipeline using ATLAS and CMS data. The pipeline achieves tracking efficiency and purity similar to production tracking algorithms. Crucially for future HEP applications, the pipeline benefits significantly from GPU acceleration, and its computational requirements scale close to linearly with the number of particles in the event.  
}

\section{Introduction
\label{sec-challenges}
\label{intro}
}
   Charged particle tracking plays an essential role in High-Energy Physics (HEP), including particle identification and kinematics, vertex finding, lepton reconstruction, and flavor jet tagging.   At the core of particle tracking there is a pattern recognition algorithm that must associate a list of 2D or 3D position measurements from a tracking detector (known as {\it hits} or {\it spacepoints} in literature) to a list of particle track candidates (or {\it tracks}. A {\it track} is defined as a list of spacepoints associated by the pattern recognition to a charged particle). 

  The number of particle track candidates varies significantly from one experiment setup to another. For example, in a High-Luminosity LHC (HL-LHC)~\cite{hllhc} collision \textit{event}, due to  the \textit{pile-up} of multiple proton-proton collision per bunch crossing,  there are typically 5,000 charged particles and 100,000 spacepoints, about 50~\% of which are associated to particles of interest. 
 
 \begin{figure}[!htb]
    \centering
    \includegraphics[width=0.9\textwidth]{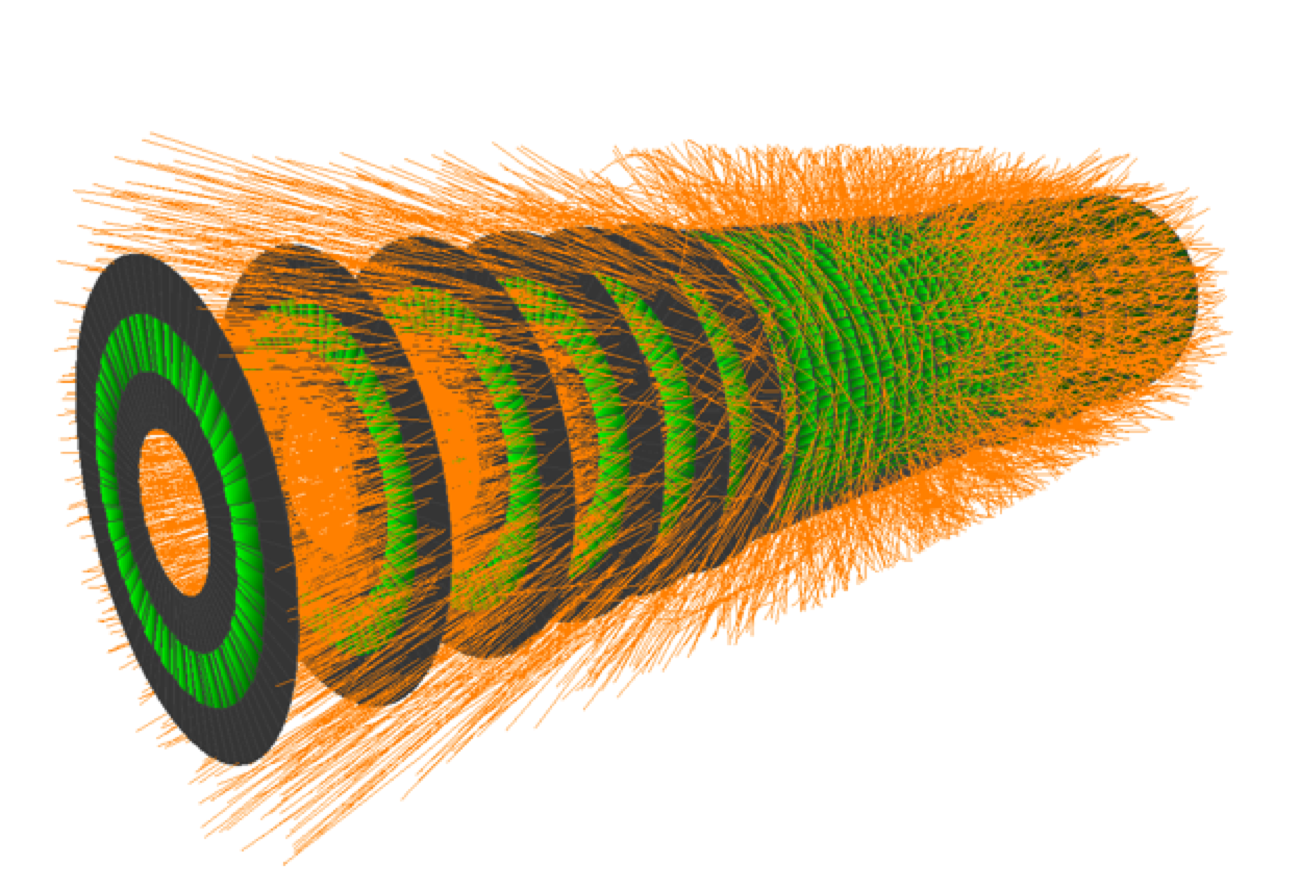} 
    \includegraphics[width=0.9\textwidth]{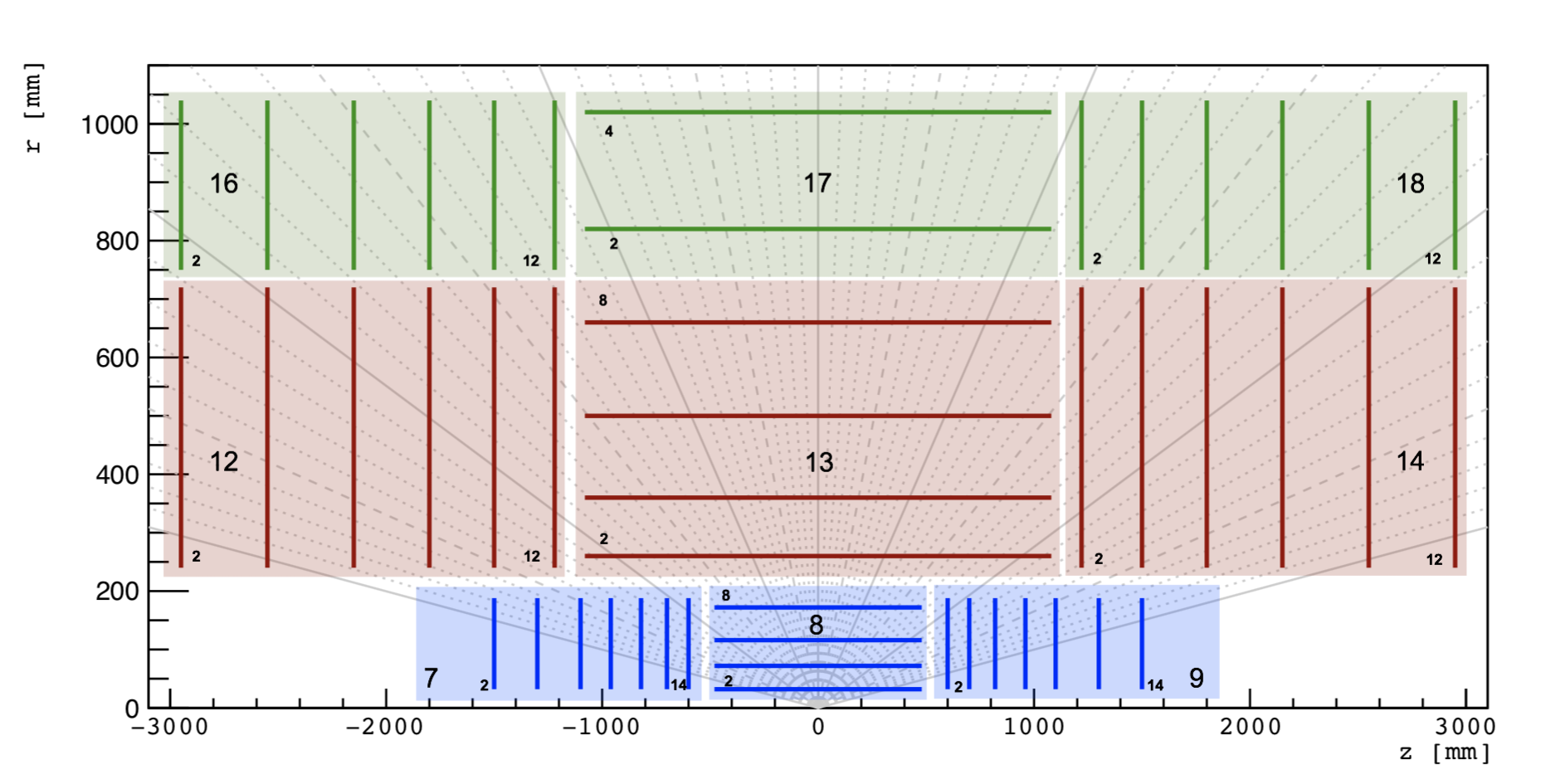}
    \caption{A simulated HL-LHC collision event (top) as seen by the TrackML tracking detector~\cite{TrackMLAccuracy2019}. The detector schematic (bottom) shows the top half of the detector projected on the r-z plane. The z-axis is along the beam direction.}
        \label{fig:trackml_detector}
\end{figure}

 A typical HEP offline tracking algorithm~\cite{RevModPhys.82.1419,TIDE,Chatrchyan:2014fea} has four stages: spacepoint formation, track seeding, track following, and track fitting.   The spacepoint formation stage combines the detector readout cell raw data in clusters from which the spacepoint 3D coordinates, and their uncertainties, are determined.  Track seeding combines spacepoints in {\it doublet} or {\it triplet} {\it seeds}. Each seed provides an initial track direction, origin, and possibly a curvature, with associated uncertainties. The track following stage adds more spacepoints to the seed by looking for matching spacepoints along the extrapolated trajectory. Finally a track fitting stage, which may be combined with the track following, fits a trajectory through the track spacepoints to assess the track quality and measure the particle's physical and kinematic properties (charge, momentum, origin, etc). To avoid biasing physics results, each stage of the algorithm must have high {\it efficiency}, meaning it must identify e.g. $> 90\%$ of the charged particles within a fiducial region (e.g. $\pt > 1$ GeV, $|\eta| < 4$) as track candidates. Track seeding and track filtering must also have high {\it purity}, meaning that e.g. $>60\%$ of the track seeds and track candidates must correspond to charged particles. High purity allows to keep the number of track candidates, and the associated computational costs, under control.
 
  
 Online tracking algorithms may use different pattern recognition algorithms\footnote{including Hough transforms~\cite{hough_transform,Gradin:2017jxr} and cellular automata~\cite{Funke:2014dga,Rohr:2019ava}} to create and filter track seeds and candidates, but share the same high efficiency requirements. Online application also have stringent computing requirements (e.g. latency $O(10)~\mu$s for LHC triggers).
 
\begin{figure}[!htb]
        \includegraphics[width=0.9\textwidth]{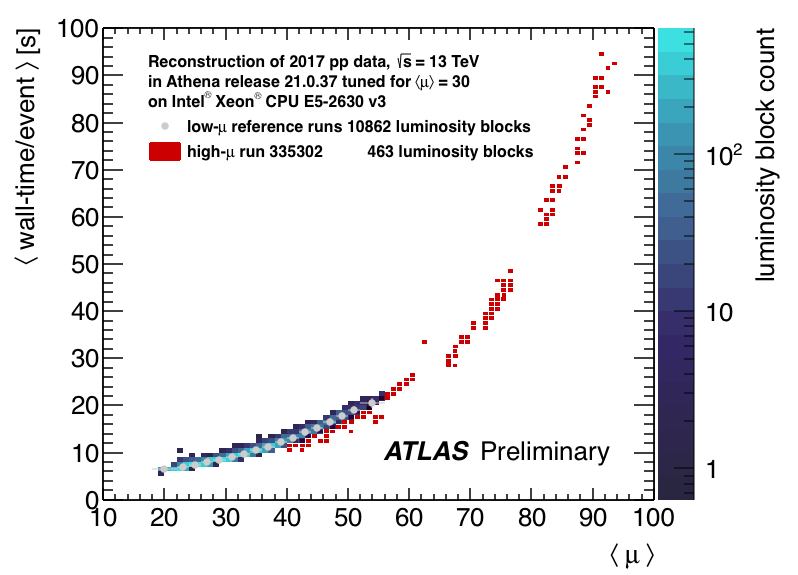} 
        \includegraphics[width=0.8\textwidth]{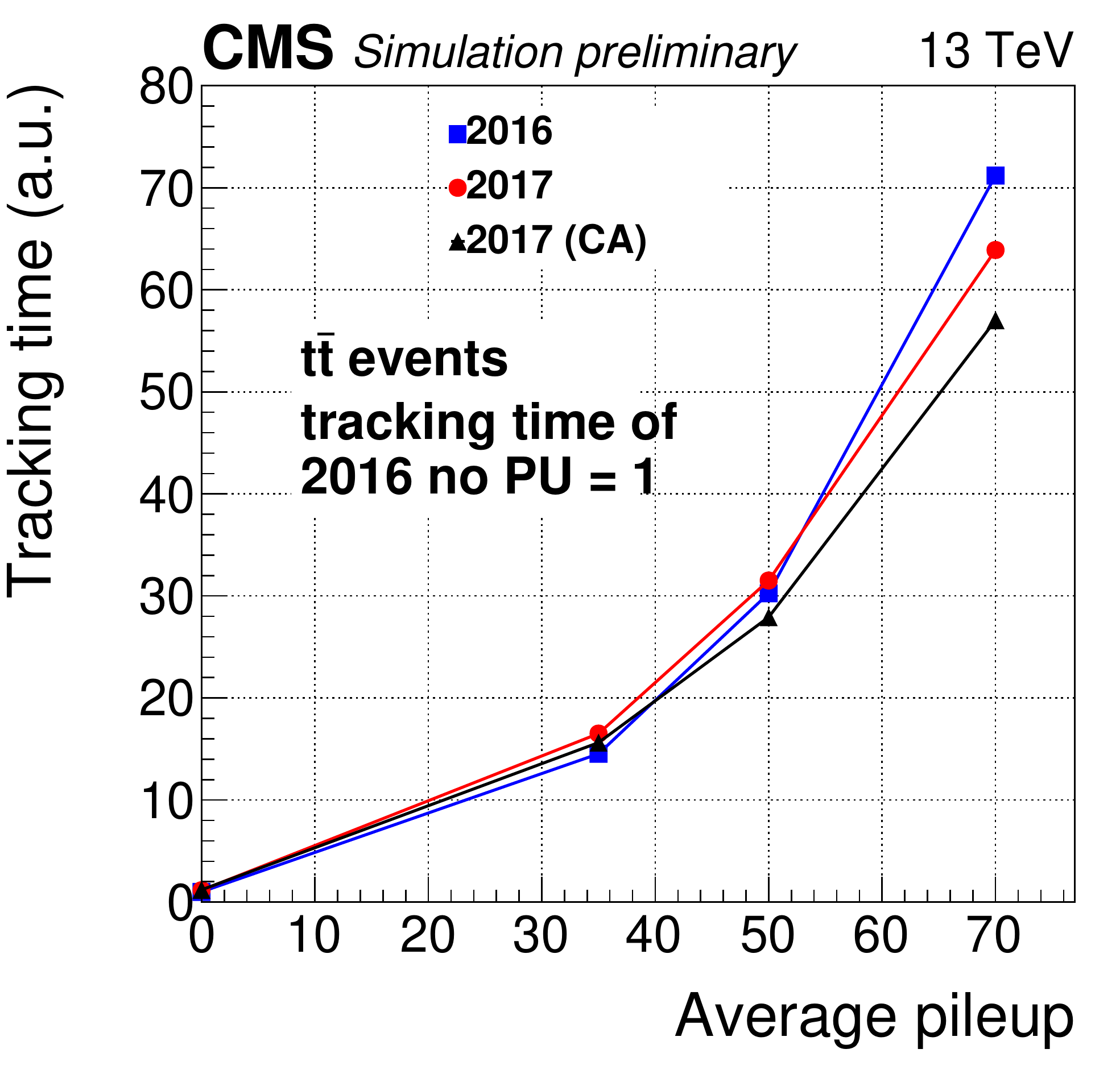} \\

    \caption{Reconstruction wall time per event as a function of the average number of interactions per bunch crossing \mmu. Top: ATLAS Run~2 Inner Detector reconstruction with default configurations~\cite{ATLASCompRes}. Bottom: CMS time spent in tracking sequence for 2016 tracking, 2017 tracking with conventional seeding, and 2017 tracking with Cellular Automaton (CA) seeding~\cite{CMSCompRes}.
    }
    \label{fig:atlas_reco_time}
\end{figure}

 The computational cost of current tracking algorithms grows worse than linearly with beam intensity and detector occupancy, as demonstrated in Figure~\ref{fig:atlas_reco_time}. Given the order-of-magnitude increase for beam intensity at HL-LHC, charged particle pattern recognition algorithms might well limit the discovery potential of HL-LHC experiments.

Over the last two decades, tracking computational challenges arising from the increased number of combinations have been addressed by tightening fiducial regions for charged particles, developing highly optimized tracking algorithms~\cite{TIDE,Chatrchyan:2014fea}, and even optimizing the geometry of tracking detectors. These optimizations brought order-of-magnitude gains in tracking computational performance with limited impact on physics. While these efforts continue~\cite{ATLAS_SW_TRIG}, it is unlikely that another order of magnitude can be gained through incremental optimization without impacting physics performance. Furthermore, given the computational complexity and iterative nature of current track following and filtering algorithms, it is challenging to run them efficiently on data parallel architectures like GPUs.

\label{sec-previous}
The TrackML challenge~\cite{TrackMLAccuracy2019} jump-started the application  of deep learning pattern recognition methods applied to HEP tracking. The HEP.TrkX pilot project~\cite{heptrkx} proposed the use of graph networks to filter track doublet and triplet seeds~\cite{heptrkx-ctd2018}. Building on that work, the Exa.TrkX project~\cite{exatrkx} has demonstrated the applicability of \textit{Geometric Deep Learning} (GDL) methods~\cite{GDL} -- specifically metric learning and Graph Neural Networks (GNN) -- to particle tracking~\cite{choma2020track}. GDL is concerned with learning representations of data that have complex geometrical relationships and no natural ordering, like detector spacepoints. GDL models are computationally regular, naturally parallel and therefore well-suited to run on hardware accelerators. 

This work describes new developments that enabled the first study of the computing and physics performance of the Exa.TrkX pipeline on the entire TrackML detector at HL-LHC design luminosity, a  step towards the validation of the pipeline on ATLAS and CMS data.

\section{Related work}
\label{sec:related}
Early on, the Hep.TrkX pilot project attempted to assign and regress track parameters to single spacepoints using image processing models. Subsequent attempts at estimating track parameters using image processing and recurrent networks showed promising results~\cite{heptrkx-ctd2017} in a simplified environment. A similar realization of the method is reported in~\cite{Bertacchi:2019wbp} where a model processing image from successive pixel detector layers is used to produce tracklets, seeds to classical pattern recognition. The method yields superior seeding efficiency for tracks within jets in dense environments.
The concept of using LSTM~\cite{LSTM} to supplement the Kalman Filter method for track following developed by HEP.TrkX~\cite{heptrkx-ctd2017,Tsaris_2018,heptrkx-ctd2018} was later found in one of the promising solutions of the accuracy phase~\cite{Amrouche:2019wmx} of the TrackML challenge.
The task of particle tracking was addressed with a hit-to-track assignment method using gated recurrent unit~\cite{cho2014learning} (GRU), producing promising result in sparse environments~\cite{Tsaris_2018}. This approach was constrained computationally due to the use of recurrent models.



Ref.~\cite{biscarat2021realistic} applies the track finding approach developed in  Ref.~\cite{Ju:2020xty} to the whole detector by exploiting a new data-driven graph construction method and large model support in Tensorflow~\cite{le2019tflms}.
Ref.~\cite{Pata:2021oez} applies a similar GNN model to the task of particle-flow reconstruction. The model has a classification objective, followed by a partial regression of generator-level particle candidate kinematics. The method performs at least as well as a classical particle-flow algorithm in HL-LHC-like collision conditions.
As part of the Exa.TrkX project, graph networks are used for LArTPC track reconstruction~\cite{hewes2021graph}. Ref.~\cite{Heintz:2020soy} explores the opportunity to implement Exa.TrkX-inspired graph networks on FPGAs.
Starting from the input stage of the Exa.TrkX pipeline, Ref.~\cite{fox20204d} studies the impact of cluster shape information on track seeding performance. In Ref.~\cite{amrouche2021hashing}, metric learning is used to improve the purity in spacepoints buckets formed using similarity hashing.
With the advent of quantum computer of increasing size came the development of quantum machine learning techniques, also applied in particle physics~\cite{Guan:2020bdl}.
In particular, inspired by the use of GNN for charged particle tracking of the Exa.TrkX team, quantum graph networks have been tested on the same problem~\cite{Tuysuz:2020eaa}. 

\section{Methodology}

\subsection{Input Data}
This study is based on the TrackML dataset that uses a Montecarlo simulation of top quark pair production from proton-proton collisions at the HL-LHC. To simulate the effect of event pileup and produce realistic detector occupancy, a Poisson random number (with $\mu=200$) of QCD "minimum bias" events are overlaid on top of the $t\bar{t}$ collisions.

\begin{figure*}
    \centering
    \includegraphics[width=0.9\linewidth]{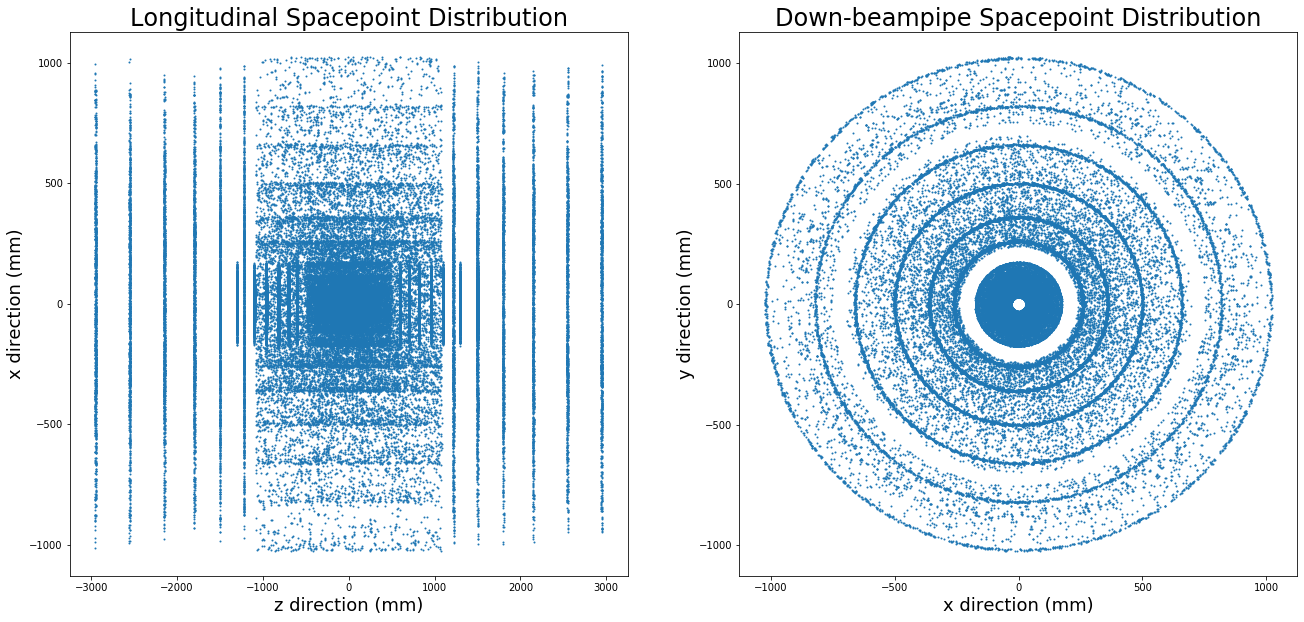}
    \caption{A typical event distribution of spacepoints projected on the $x$-$z$ plane, parallel to the beam direction (left), and the $x$-$y$ plane, orthogonal to the beam direction (right).}
    \label{fig:distributions}
\end{figure*}

The TrackML detector is a set of concentric cylindrical layers of pixelated sensors (the \textit{barrel}) complemented by a set of circular disks (the \textit{endcaps}) to ensure nearly $4\pi$ coverage in solid angle, as pictured in Figure~\ref{fig:trackml_detector}. Figure~\ref{fig:distributions} shows the spatial distribution of the spacepoints of a typical event. 
One notable feature of this dataset is the inclusion of ``noise'' spacepoints, added as a proxy for various low-momentum particle interactions and detector effects which would otherwise require more expensive and detailed simulations.

\subsection{The Geometric Deep Learning Pipeline}
\begin{figure*}
    \centering
    \includegraphics[width=0.9\linewidth]{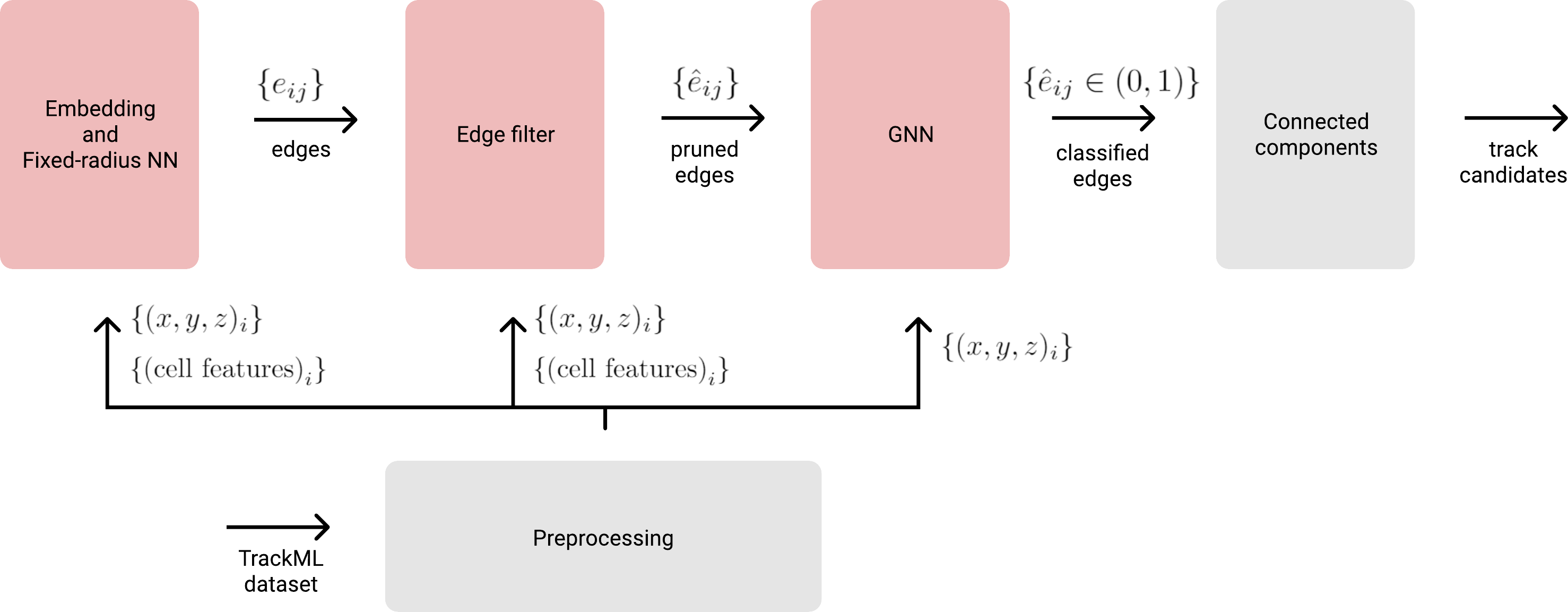}
    \caption{Stages of the TrackML track formation inference pipeline. Light red boxes are trainable stages. }
    \label{fig:pipeline}
\end{figure*}

This paper updates the methodology previously presented in Ref.~\cite{choma2020track} to a fully-learned pipeline, where both graph construction and graph classification are trained. This section describes the pipeline (represented schematically in Figure~\ref{fig:pipeline}) used to obtain the results in \S~\ref{sec:results}. Details of the latest model design, parameter choices, and technical optimizations are discussed in \S~\ref{sec:discussion}.

The pipeline currently used to reconstruct tracks from a pointcloud of spacepoints requires six discrete stages of processing and inference. These broadly consist of a preprocessing stage, three stages required to construct a spacepoint graph, and two stages required to classify the graph edges and partition them into track candidates. Each stage is trained independently (due to memory constraints) on the output of the previous stage's inference.

First, the dataset is processed into a format suitable for model training. This includes calculating directional information and summary statistics from the charge deposited in each spacepoint, i.e. the \textit{cell features} in Figure~\ref{fig:pipeline}. These values are appended to the cylindrical coordinates of each spacepoint to form an input feature vector to the pipeline. To apply a graph neural network to this set of data, it is necessary to arrange them into a graph. One can apply various geometric heuristics to define which spacepoints are likely to be connected by an edge (i.e. belong to the same track), but a useful technique is to train a model on the geometry of connected tracks. Thus, our second stage is to train an Embedding Network -- a multi-layer perceptron (MLP) which embeds each spacepoint into an N-dimensional latent space. The graph is constructed by connecting neighboring spacepoints within a radius $r_{\textnormal{embedding}}$, in the latent space. We train this embedding with a pairwise hinge loss, to encourage spacepoints that belong to the same track to be close in the embedded space, according to the Euclidean metric. This allows for a highly efficient edge construction, since we do not rely on any heuristics of the detector geometry that may lead to missed edges. 

The edge selection at this stage is close to 100\% efficient but $O(1)\%$ pure, with a graph size of $O(10^5)$ nodes and $O(10^7)$ edges (the purity-efficiency trade-off can be tuned with the choice of $r_{\textnormal{embedding}}$). Before running training or inference on the memory-intensive GNN, we filter these edges down with another MLP. The input to this third stage is the concatenated features on either side of each edge. That is, the Filter Network is a binary classifier applied to the set of edges. Constraining edge efficiency to remain high (above 96\%) leads to much sparser graphs, of $O(10^6)$ edges. 

The fourth stage of the pipeline is the training and inference of the graph neural network. The results presented in this work are predominantly obtained from the Interaction Network architecture, first proposed in Ref.~\cite{interaction-networks}. This varietal of GNN includes hidden features on both nodes and edges, which are propagated around the graph (called “message passing”) with consecutive concatenations along edges and aggregations of messages at receiving nodes. In the final layer of the network, a binary classification is obtained for each edge as true or fake, and trained on a cross-entropy loss. 

The final stage of the TrackML pipeline involves task-specific post-processing. If our goal is track formation, we can place a threshold on the edge scores produced by the GNN and partition the graph into connected components. If our goal is track seeding, we can directly sample the classified edges for high likelihood combinations of connected triplets, or convert the entire graph to a \textit{triplet graph} and train this on a second GNN to classify the triplets. A triplet graph is formed by taking all edges in the original (\textit{doublet}) graph and assigning them as nodes in the new triplet graph. The nodes in this triplet graph are connected if they share a hit in the doublet graph. Applying a GNN to this structure produces highly pure sets of seeds as shown in Ref.~\cite{choma2020track}.

Many of these techniques are common to other applications being explored in the Exa.TrkX collaboration. The pattern of nearest-neighbor graph-building and  GNN edge classification has shown its potential for neutrino experiments~\cite{hewes2021graph} and CMS High Granularity Calorimeter~\cite{Ju:2020xty}. 
Indeed, these applications build on the TrackML pipeline and extend it, for example by adding the particle type as an edge feature.

\section{Results}
\label{sec:results}
\subsection{Tracking Performance of the TrackML pipeline}
\subsubsection{Tracking Efficiency and Purity}
\label{sec:effpur}
The performance of a tracking pipeline is mainly characterized by tracking efficiency and purity. For efficiency calculations, only charged particles that satisfy 
$|\eta| < 4.0$ and $\pt > 100$~MeV are considered. These \textit{selected} particles, $N_{particles}(\textnormal{selected})$, are hereafter referred to as \textit{particles}.

The overall tracking efficiency, known as \textit{physics efficiency} ~$\epsilon_\textnormal{phys}$~(Eq.~\ref{eqn:effp}), is defined as the fraction of particles that are \textit{matched} to at least one reconstructed track.  A particle is considered to be matched to a reconstructed track when 1) the majority of spacepoints in the reconstructed track belong to the same true track, and 2) the majority of spacepoints in the matched true particle track are found in the reconstructed track\footnote{This nomenclature and the associated definitions broadly follow ~\cite{CERN-LHCC-2017-021,TrackMLAccuracy2019}.}.

To measure the efficiency of the tracking pipeline itself, we also define the \textit{technical efficiency}~$\epsilon_\textnormal{tech}$~(Eq.~\ref{eqn:efft}) as the fraction of \textit{reconstructable} particles matching at least one reconstructed track. Reconstructable particles have a trajectory that leaves at least five spacepoints in the detector. Tracking purity~(Eq.~\ref{eqn:purity}) is defined as the fraction of reconstructed tracks that match a selected particle\footnote{HEP tracking literature often quotes $
\textnormal{fake rate} = 1 - \textnormal{purity}$}.

\begin{align}
 \epsilon_\textnormal{phys} &= \frac{N_{particles}(\textnormal{selected, matched})}{N_{particles}(\textnormal{selected})} \label{eqn:effp}\\
\epsilon_\textnormal{tech} &= \frac{N_{particles}(\textnormal{selected, reconstructable, matched})}{N_{particles}(\textnormal{selected, reconstructable})} \label{eqn:efft}
\end{align}

\begin{equation}
\textnormal{Purity} = 
\frac{N_{tracks}(\textnormal{selected,matched})}
     {N_{tracks}(\textnormal{selected})} \label{eqn:purity}
\end{equation}

Averaged over 50 testing events from the TrackML dataset, the physics efficiency for particles with $\pt > 500$~MeV is $88.7\pm 0.3\%$ and the technical efficiency is $97.6\pm 0.3\%$. Without any fiducial \pt~cut, the physics efficiency becomes $67.2\pm 0.1\%$ and the technical efficiency $91.3\pm 0.2\%$. The tracking purity is $58.3\pm 0.6\%$. Using the TrackML challenge scoring system and all tracks in the event, we obtained a score of $0.877\pm 0.005$~\footnote{We obtained a score of $0.914\pm 0.006$ by training the pipeline with a dataset that includes noise hits, that we otherwise removed from our training dataset to facilitate the noise impact studies of section 4.1.2}. The errors quoted are statistical only.

\begin{figure*}
    \centering
    \includegraphics[width=0.45\textwidth]{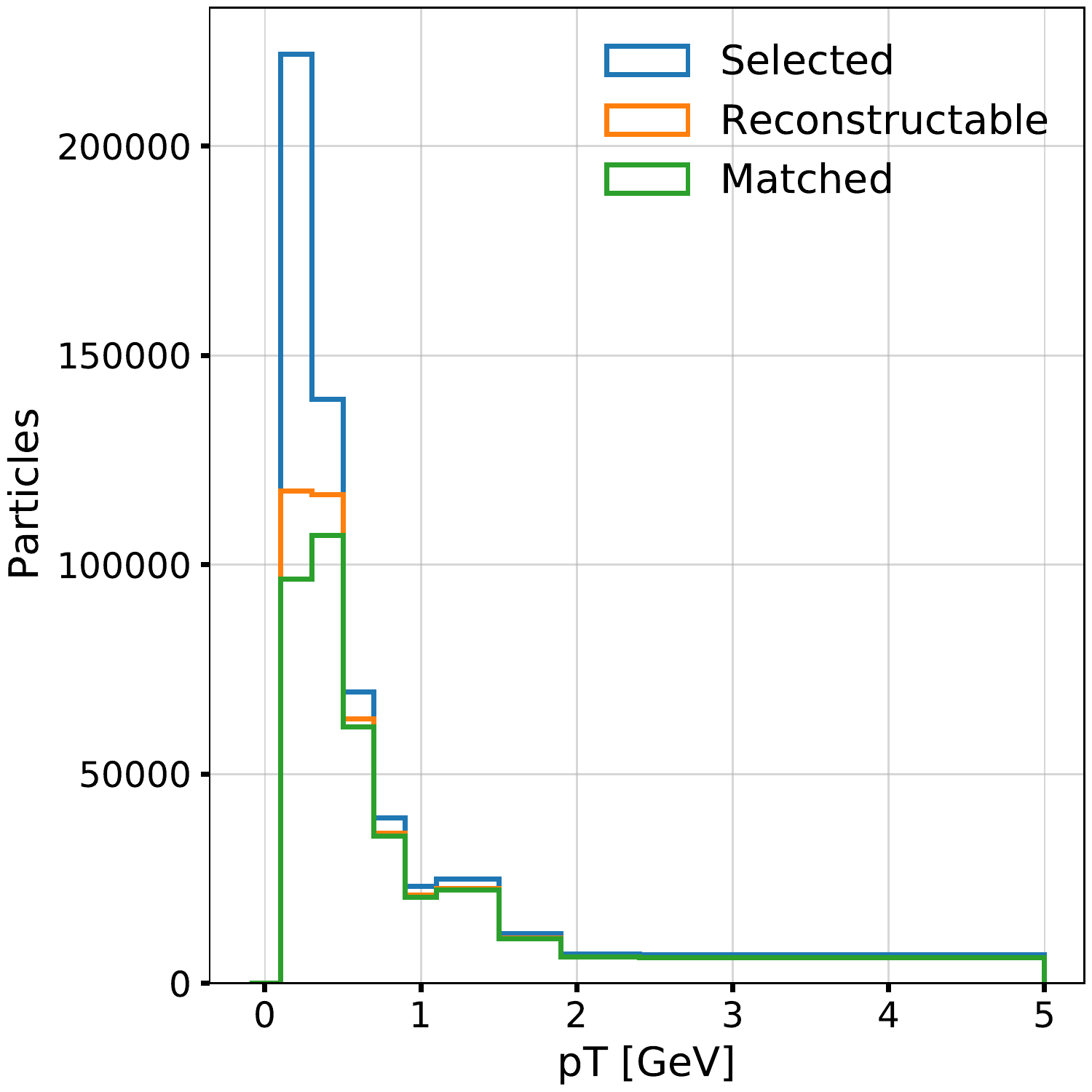}
    \includegraphics[width=0.45\textwidth]{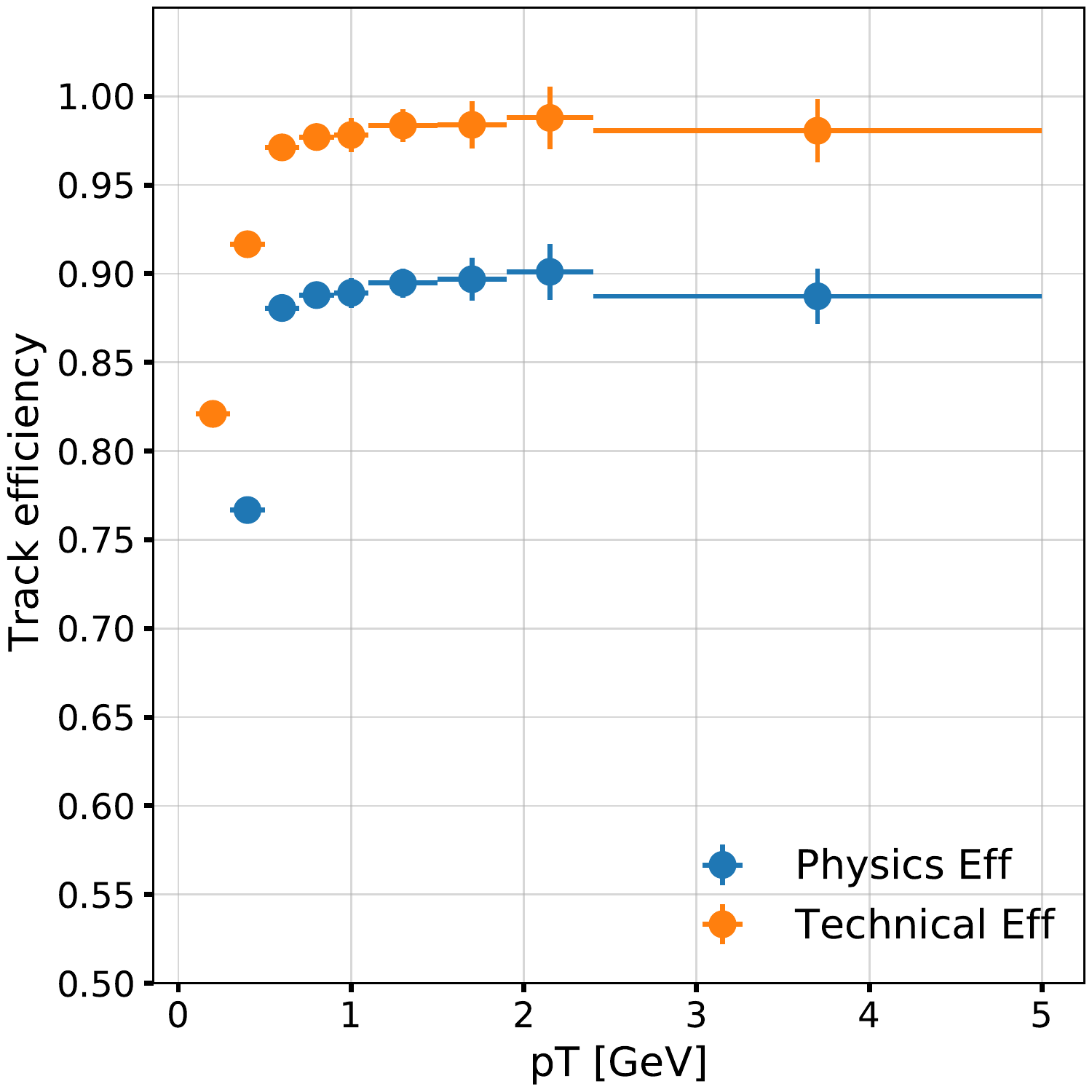}
    \includegraphics[width=0.45\textwidth]{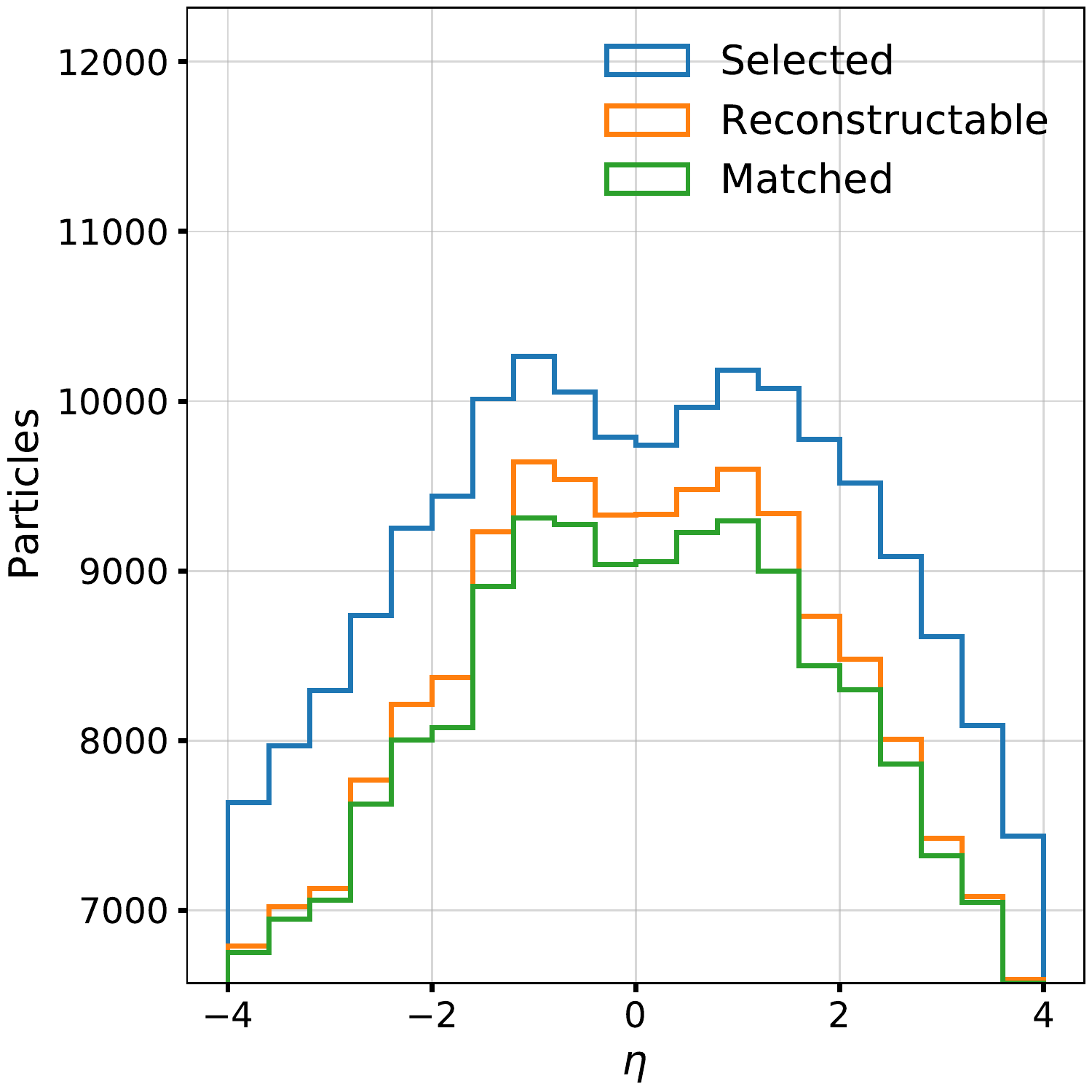}
    \includegraphics[width=0.45\textwidth]{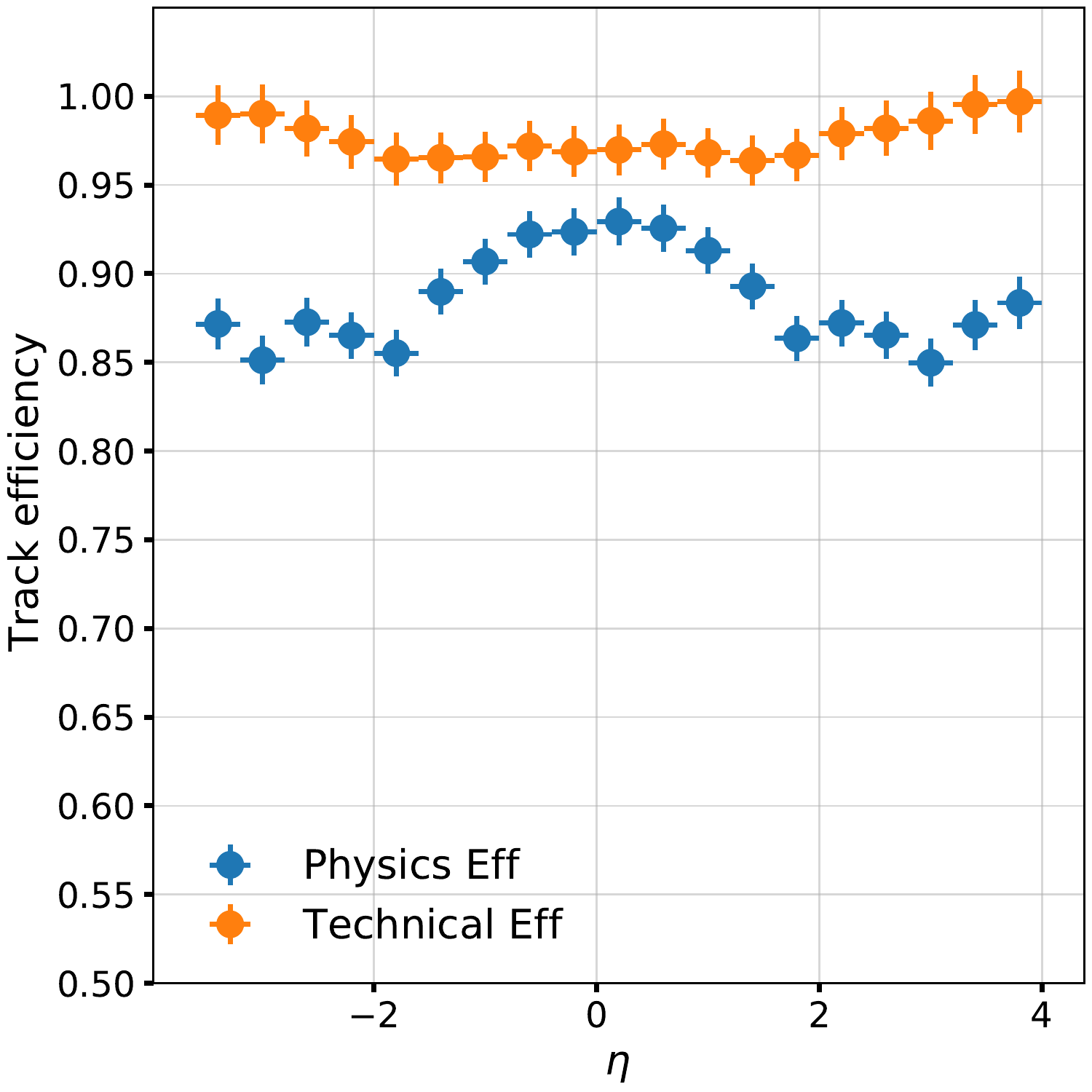}
    \caption{
    Top row: selected, reconstructable, and matched particles (left) and tracking efficiency (right) as a function of \pt\ for particles with $|\eta| < 4$. Bottom row: selected, reconstructable, and matched particles (left) and tracking efficiency (right) as a function of $\eta$ for $\pt > 0.5$~GeV. The definition of ``selected'', ``reconstructable'', and ``matched'' can be found in \S~\ref{sec:effpur}
        \label{fig:res_pt_eta}
    }

\end{figure*}

Figure~\ref{fig:res_pt_eta} shows the \pt\ distribution of particles as well as the tracking efficiency as a function of particle \pt. The physics efficiency for particles with \pt\ of [100, 300]~MeV is 43\%, therefore, is not displayed in the plot. The physics efficiency for particles with $\pt > 700$~MeV is above 88\%. The technical efficiency is 82\% for particles with \pt\ of [100, 300]~MeV, and increases to above 97\% for particles with $\pt > 700$~MeV. 
 Figure~\ref{fig:res_pt_eta} also shows the $\eta$ distribution of particles with $\pt > 500$~MeV as well as the tracking efficiency as a function of the particle $\eta$. The physics efficiency is higher in the \textit{barrel} region of the detector (volumes 8,13,17 in Figure~\ref{fig:trackml_detector}), while the technical efficiency is almost flat across the $\eta$ range.
In Figure~\ref{fig:res_pt_eta} the \pt\ and $\eta$ of the matched truth particle were used, rather than the \pt\ and $\eta$ of the reconstructed track. We leave a study of track quality and detector resolution effects for future work.

\subsubsection{Systematic Studies}
Before using a tracking algorithm in production, it is necessary to measure its sensitivity to systematic effects, including pile-up, noise and digitization errors, and uncertainties in the measurement of detector properties (alignment, rotation, magnetic field map, etc.).

\begin{figure}[!htb]
    \includegraphics[width=0.9\textwidth]{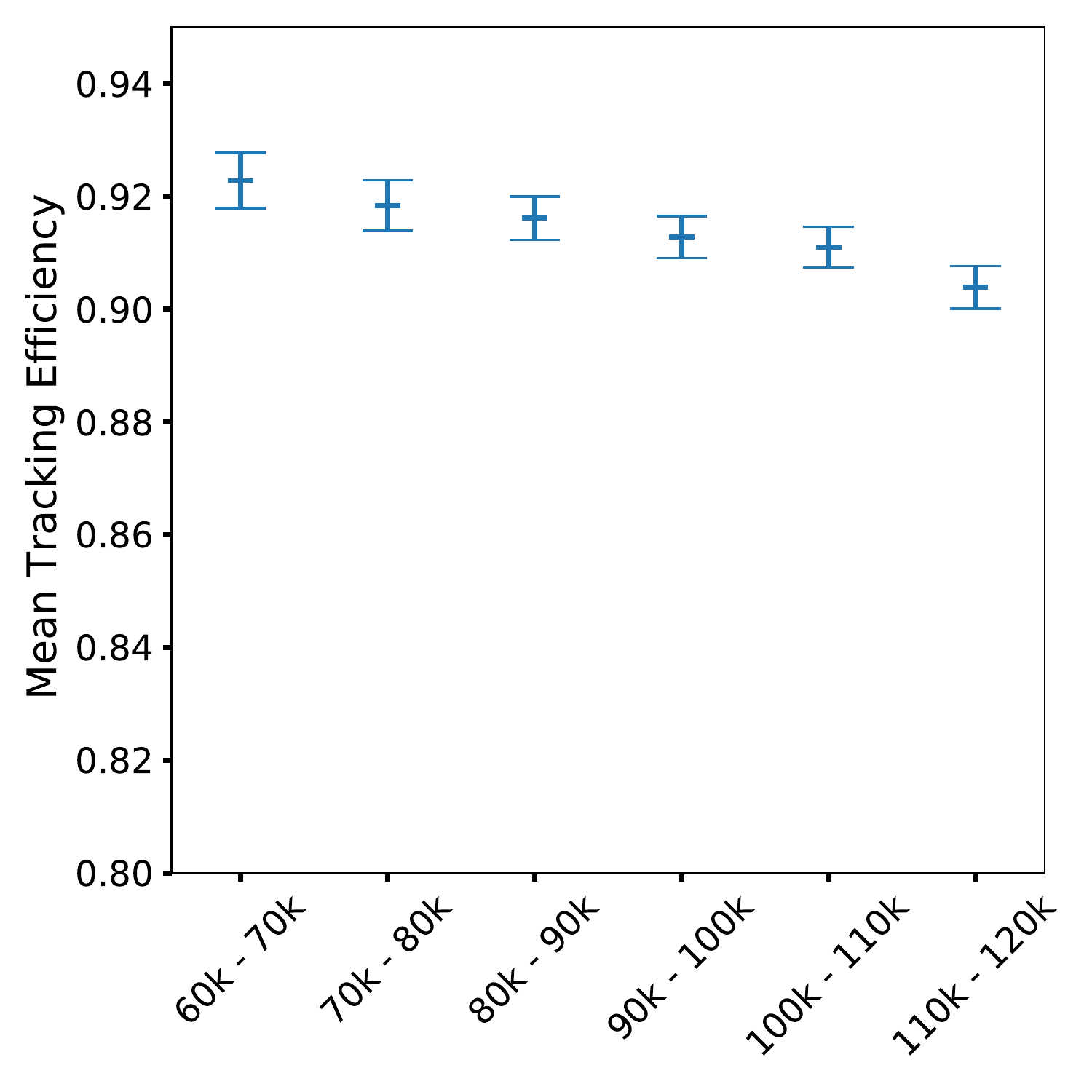}
    \includegraphics[width=0.9\textwidth]{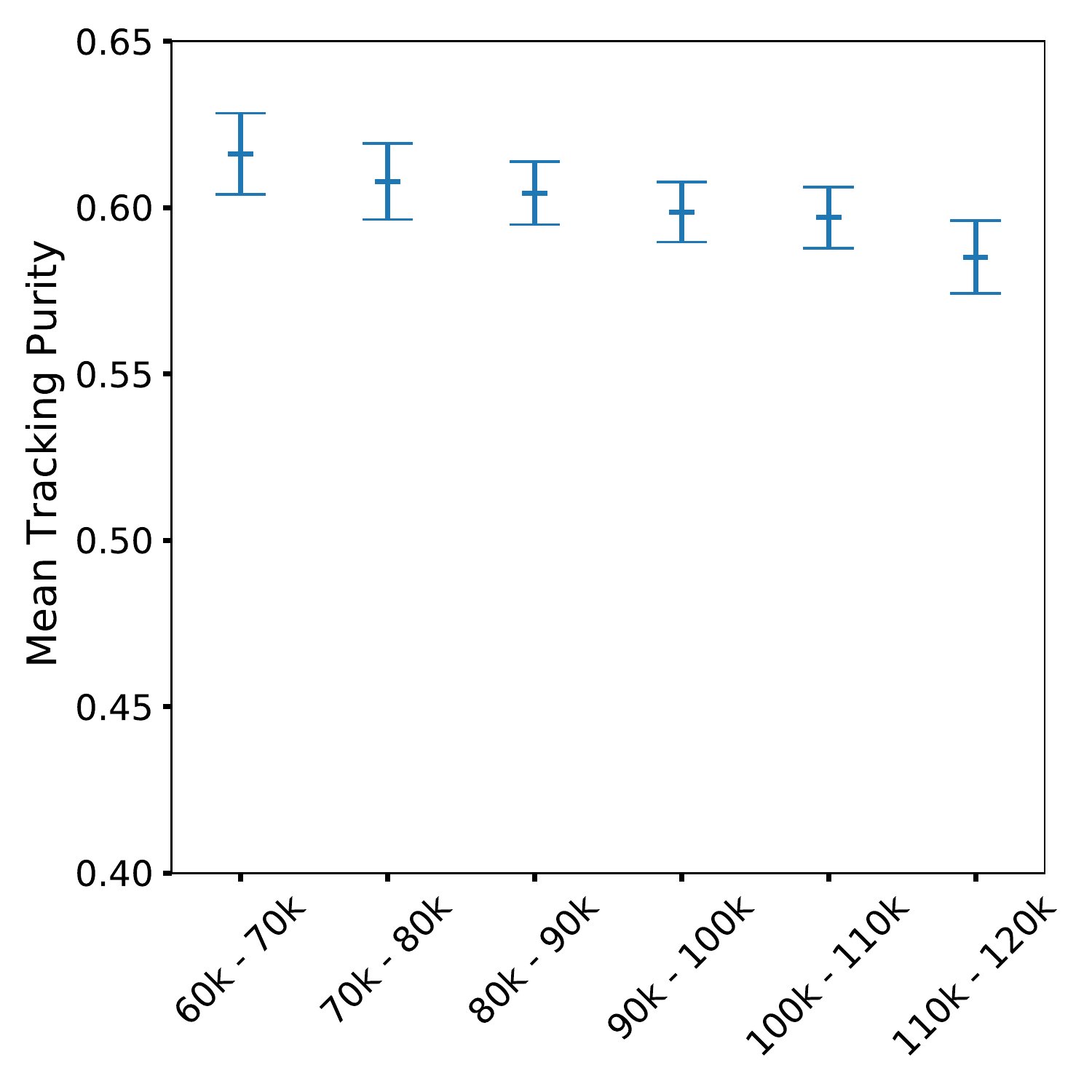}
    \caption{Mean and standard deviation of the technical efficiency (top) and purity (bottom) as a function of the total number of spacepoints in an event.}
    \label{fig:res_vs_hits}
\end{figure}

Measuring precisely the impact of  pile-up collisions on tracking performance is beyond the scope of this work, but we can estimate pile-up's impact on tracking performance by plotting efficiency and purity as a function of the number of spacepoints in the detector. Figure~\ref{fig:res_vs_hits} shows that the effect of the increased detector occupancy is a smooth performance degradation O(\%). In future work, we will study the origin of this degradation to achieve the stable performance of traditional algorithms \cite{ATLAS-TDR-030}.

\begin{figure}[!htb]
   \includegraphics[width=0.9\textwidth]{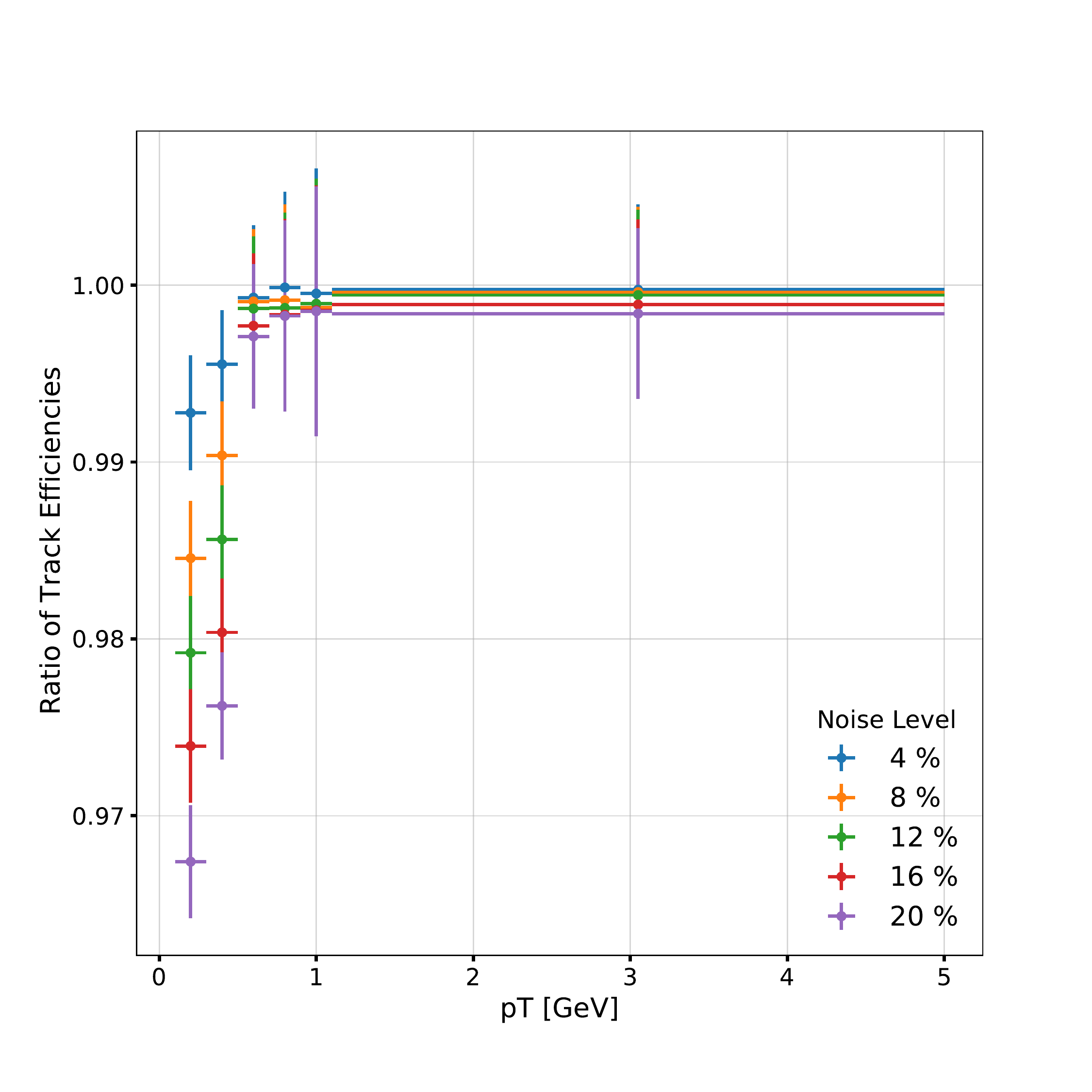}
  \caption{\textit{Relative} technical efficiency as a function of \pt. Each curve shows the ratio of $\textnormal{eff}(\textnormal{noise}=N\%) /
                                             \textnormal{eff}(\textnormal{noise}=0)$. }%
\label{fig:noise}

\end{figure}

The impact of noise spacepoints can be estimated using the TrackML dataset by studying the inference performance of the tracking pipeline, trained without any noise spacepoints, as a function of the fraction of noise spacepoints (up to a maximum of 20\% of the total). Table~\ref{tab:noise} shows the technical tracking efficiency and purity for different noise levels. The efficiency decreases by $\simeq 1.6\%$ and the purity by $\simeq 5.4\%$ when 20\% of noise spacepoints are presented. The loss of efficiency happens primarily for particles with $\pt < 500$~MeV (Figure~\ref{fig:noise}).

\begin{table}[!htb]
    \begin{tabular}{|c|c|c|}
    \toprule
    Noise & $\epsilon_\textnormal{tech}$ & Purity \\ \hline
    0       & 91.5  & 59.3 \\
    4\%     & 91.5  & 59.3  \\
    8\%     & 91.1  & 58.0 \\
    12\%    & 90.9  & 56.8 \\
    16\%    & 92.2  & 54.8 \\
    20\%    & 89.9  & 53.9 \\
    \bottomrule
  \end{tabular}

  \caption{Technical efficiency and purity for different noise fractions $(N^{\textnormal{noise}}_{\textnormal{spp}}/N_{\textnormal{spp}})*100\%$}%
      \label{tab:noise}
\end{table}

Detector misalignment effects are approximated by shifting by up to 1~mm the $x$-axis of all spacepoints in the inner-most TrackML barrel detector layer or the four innermost layers (volume 8 in Figure~\ref{fig:trackml_detector}). In both cases, the impact on the tracking efficiency is less than 0.1\%. However, studying in depth misalignments, and other detector effects, requires access to experiment detailed detector simulation data. We leave these studies as future work to be performed in collaboration with each experiment. 

\subsection{Distributed Training Performance}
\label{sec:distrmethods}

Our training sample consists of 7500 pileup events from the TrackML dataset. It takes about 1.5 days to train the Exa.TrkX pipeline on a Nvidia A100 GPU for a set of hyper-parameters. It is therefore desirable to use distributed training to parallelize model training and hyper-parameter optimization (HPO). This study relied on data parallel training~\cite{bennun2018demystifying}
implemented using  Horovod~\cite{sergeev2018horovod} and Tensorflow's \textsc{tf.distributed} framework~\cite{abadi2016tensorflow}. Horovod supports distributed training across multiple nodes, while \textsc{tf.distributed} allows to use the same code across CPUs, TPUs, and GPUs.

\begin{figure*}[!htb]
    \centering
    \includegraphics[width=0.45\textwidth]{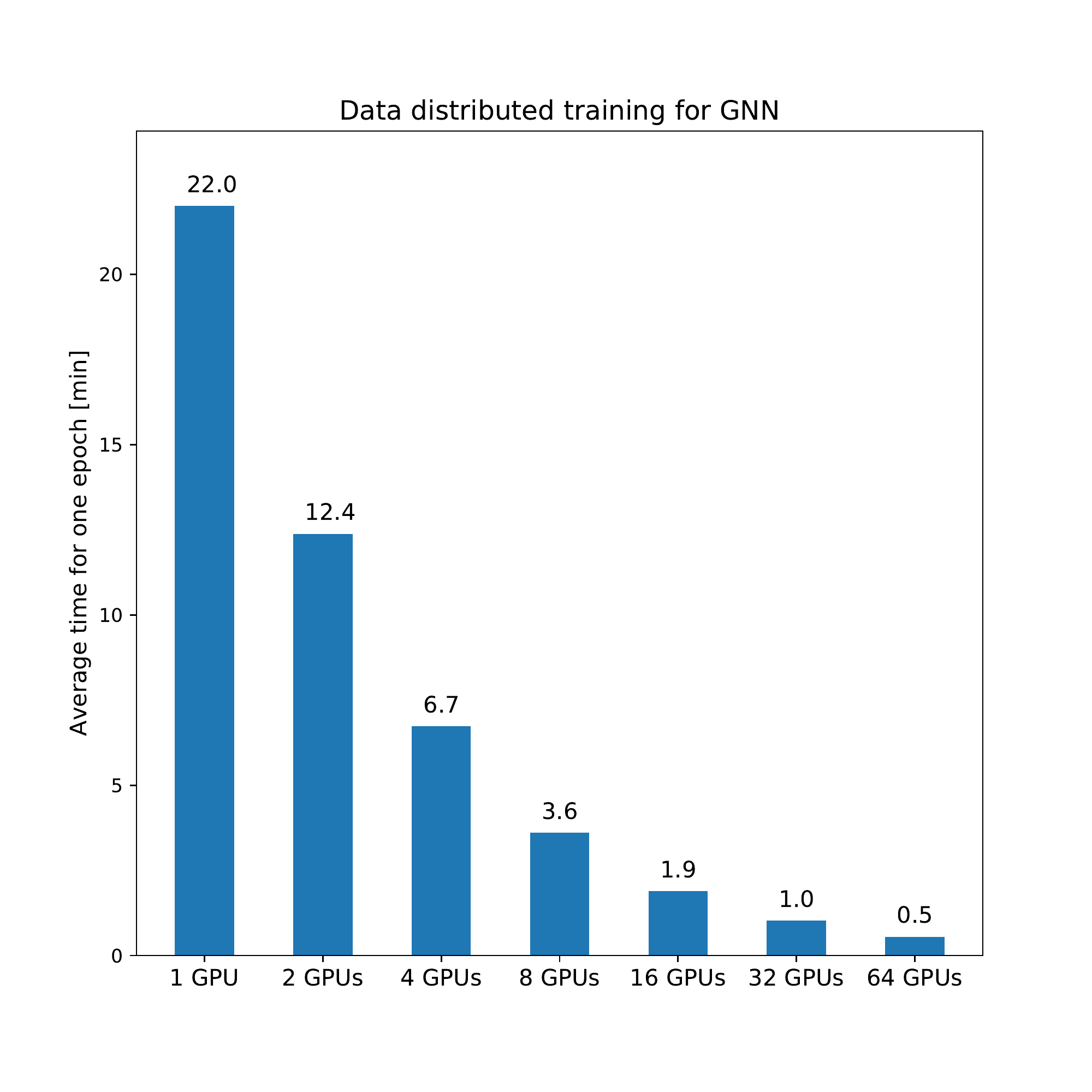}
    \includegraphics[width=0.45\textwidth]{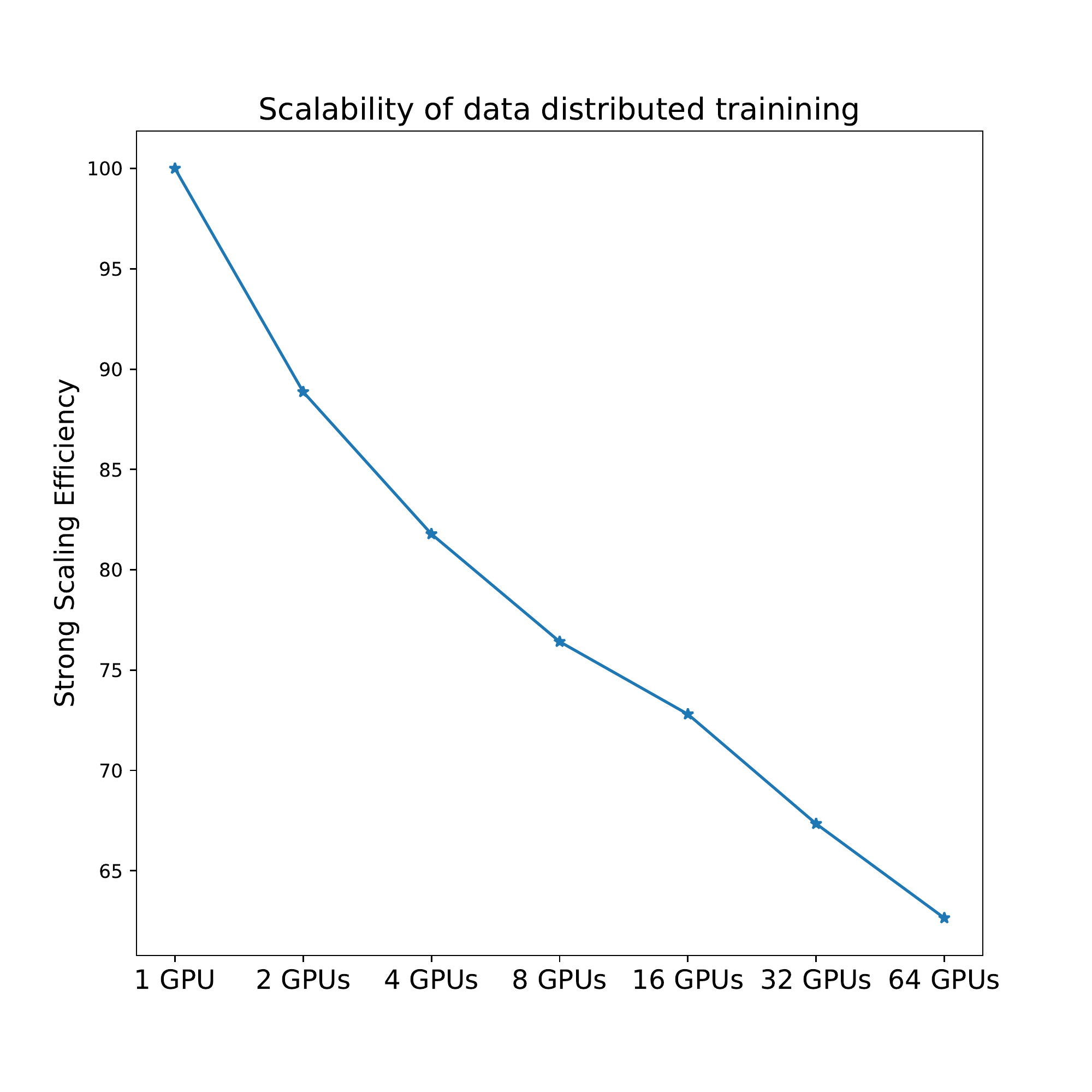} \\
    \includegraphics[width=0.45\textwidth]{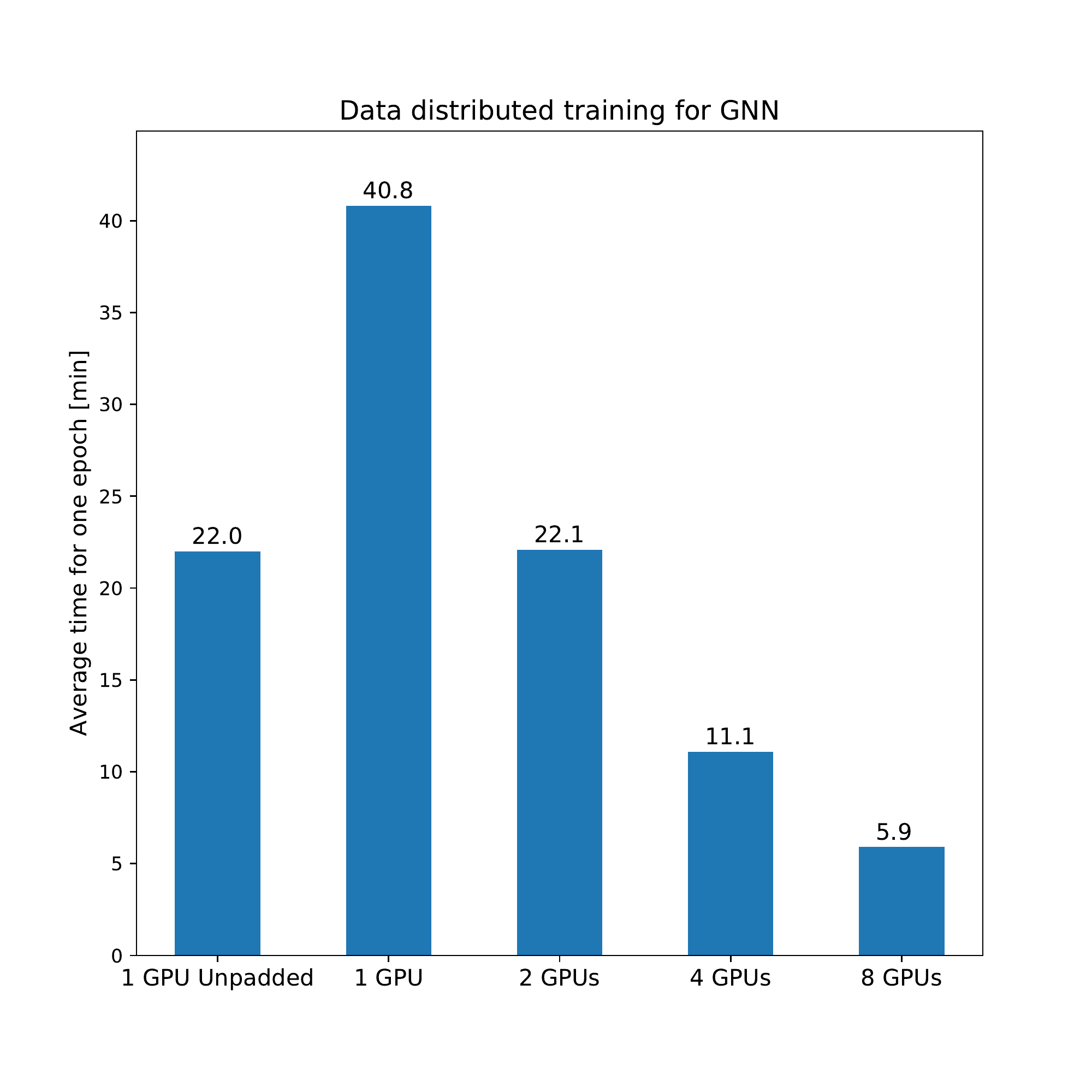}
    \includegraphics[width=0.45\textwidth]{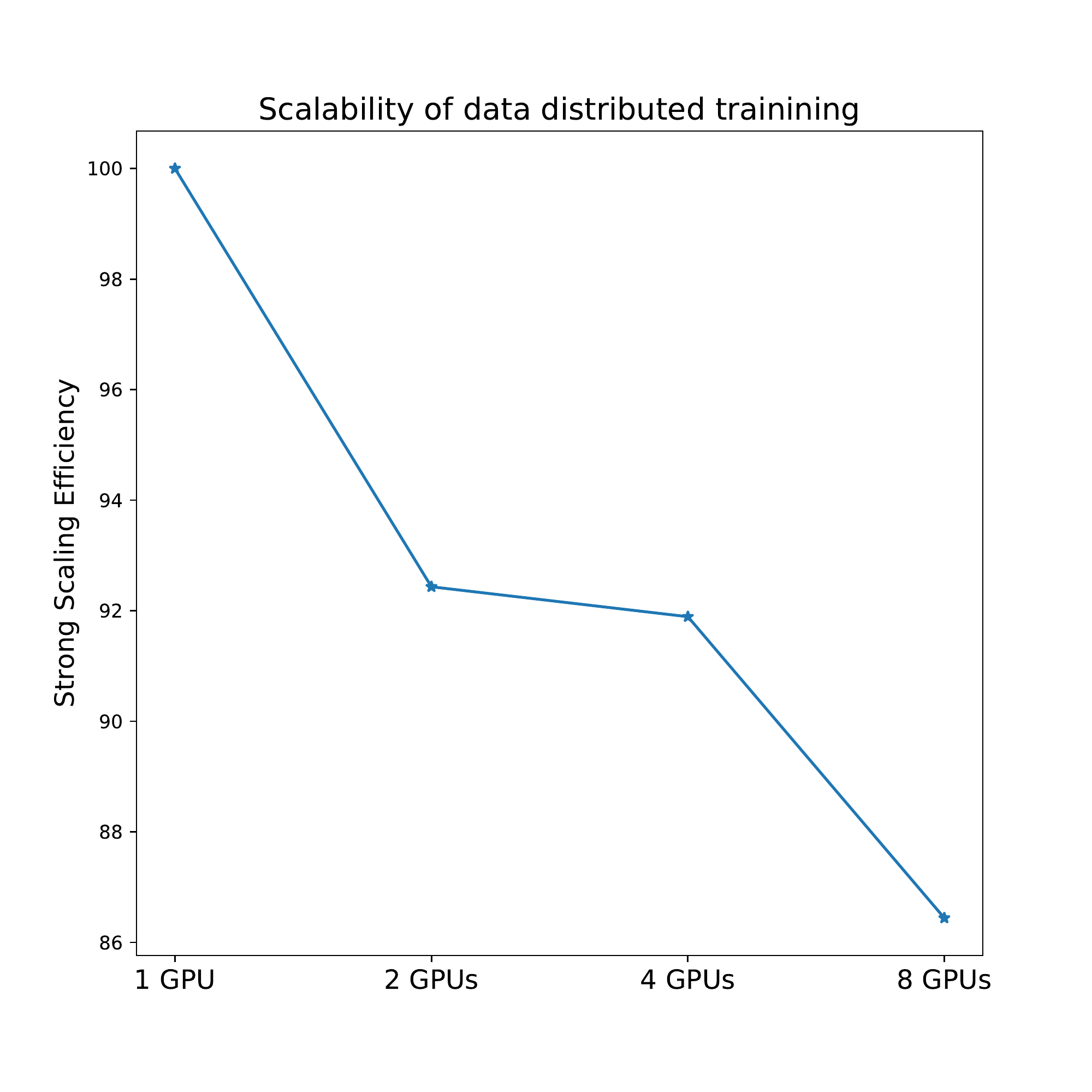}
    \caption{Time per training epoch (left) and Strong scaling efficiency (right) for GNN's distributed training. The top row refers to the Horovod implementation, the bottom row to the \textsc{tf.distributed} one. The first bin in the bottom left diagram refers to the serial case, in which the input graph is not padded.}
    \label{fig:strong_scaling_tf}
    \label{fig:strong_scaling_hvd}
\end{figure*}

For this study, the TrackML pipeline is trained on up to 64 Nvidia V100 GPUs across eight NERSC Cori-GPU computing nodes. Using the Horovod framework (Figure~\ref{fig:strong_scaling_hvd}), training time is reduced from 22 minutes, with 1 GPU, to 0.5 minutes with 64 GPUs~\footnote{All measurements in this section were taken training on spacepoints from the barrel region of the TrackML detector.  For comparison, training with spacepoints from the whole detector takes $\simeq$70 minutes per epoch on one Nvidia A100 GPU}. The strong scaling efficiency~\footnote{defined as $t_1 / (N\times t_N) * 100\%$ where $t_N$ is the time to train on a fixed total number of events across $N$ GPUs.} is about 90\% with 2 GPUs and 75\% with 8 GPUs. This deviation from ideal scaling is due to the model setup time and data movement costs. 

Figure~\ref{fig:strong_scaling_tf} also shows the scaling behaviour of the \textsc{tf.distributed} implementation. Since this implementation  requires all input data to be of the same size, we have to pad all input graphs to a fixed size. This essentially doubles the time needed to train one epoch, that increases from 22 minutes for dynamic input graph sizes to 41 minutes for constant graph sizes. Leaving aside this fixed overhead, \textsc{tf.distributed} appears to scale better than Horovod, achieving $\simeq 85\%$ strong scaling efficiency  with 8 GPUs.

\subsection{Inference performance on CPU and GPU}
It is crucial to characterize the computational cost of the end-to-end learned tracking algorithm. We rely on the \textsc{Pytorch} and \textsc{TensorFlow} libraries to optimize our inference pipeline on CPU and GPU. 
The execution time for the inference pipeline has been measured on two hardware platforms: Nvidia V100 GPUs with 16 GB on-board memory, and Intel Xeon 6148s (Skylake) CPUs with 40 cores and 192 GB memory per node. The inputs to the filtering step do not fit into the GPU memory. Therefore, edge filtering for one event is executed in mini-batches with a fixed batch size of 800k edges. Typically, the inputs to the filtering from one event are split into seven batches, leading to additional computational cost for moving data from host to GPU. The peak GPU memory consumption is about 15.7 GB as obtained from the Nvidia profiling tool. 

Averaging over 500 events, it takes $2.2 \pm 0.3$ wall-clock seconds per event (as measured by  measured by the python module \textit{time}) to run the inference pipeline on the GPU and $202 \pm 35$ seconds to run it on a single CPU core. This total execution time includes every step of the calculation, and in particular the time needed to move data from host to GPU. 
Table ~\ref{tab:inf_timing} breaks down the wall-clock time for the most significant steps of the pipeline. The results show how the graph creation and filtering steps are the biggest targets for further optimization in order to surpass traditional algorithms in terms of inference time~\cite{ATL-PHYS-PUB-2019-014}.

\begin{table}[!htb]
    \centering
    \caption{Average inference time for synchronous execution of the TrackML pipeline benchmarked on CPUs and GPUs.  For these step-by-step measurements, we force the pipeline to execute serially by calling {\tt torch.cuda.synchronize} after each step.
    The total inference time comprises all the steps including ones not listed in the table. \label{tab:inf_timing}}
    \begin{tabular}{|l|c|c|}
    \toprule
                  & Wall time [s] on & Wall time [s] on \\
                  & Xeon 6148s & Nvidia V100\\
                  & single core& synchronous\\ \hline
     Data Loading & 0.0049 $\pm$ 0.0153 &  0.0023 $\pm$ 0.0003 \\
     Embedding    & 3.02 $\pm$ 0.39 & 0.024 $\pm$ 0.003 \\
     Build Edge   & 66 $\pm$ 13 & 0.76 $\pm$ 0.10 \\
     Filtering    & 99 $\pm$ 19 & 1.57 $\pm$ 0.34 \\
     GNN          & 27 $\pm$ 2 & 0.45 $\pm$ 0.06 \\
     Labeling     & 3.23 $\pm$ 0.34 & 0.08 $\pm$ 0.01 \\
     Total (sync) & 202 $\pm$ 35 & 3.3 $\pm$ 0.5 \\
     \bottomrule
    \end{tabular}
\end{table}

\begin{figure}[!htb]
    \includegraphics[width=0.9\textwidth]{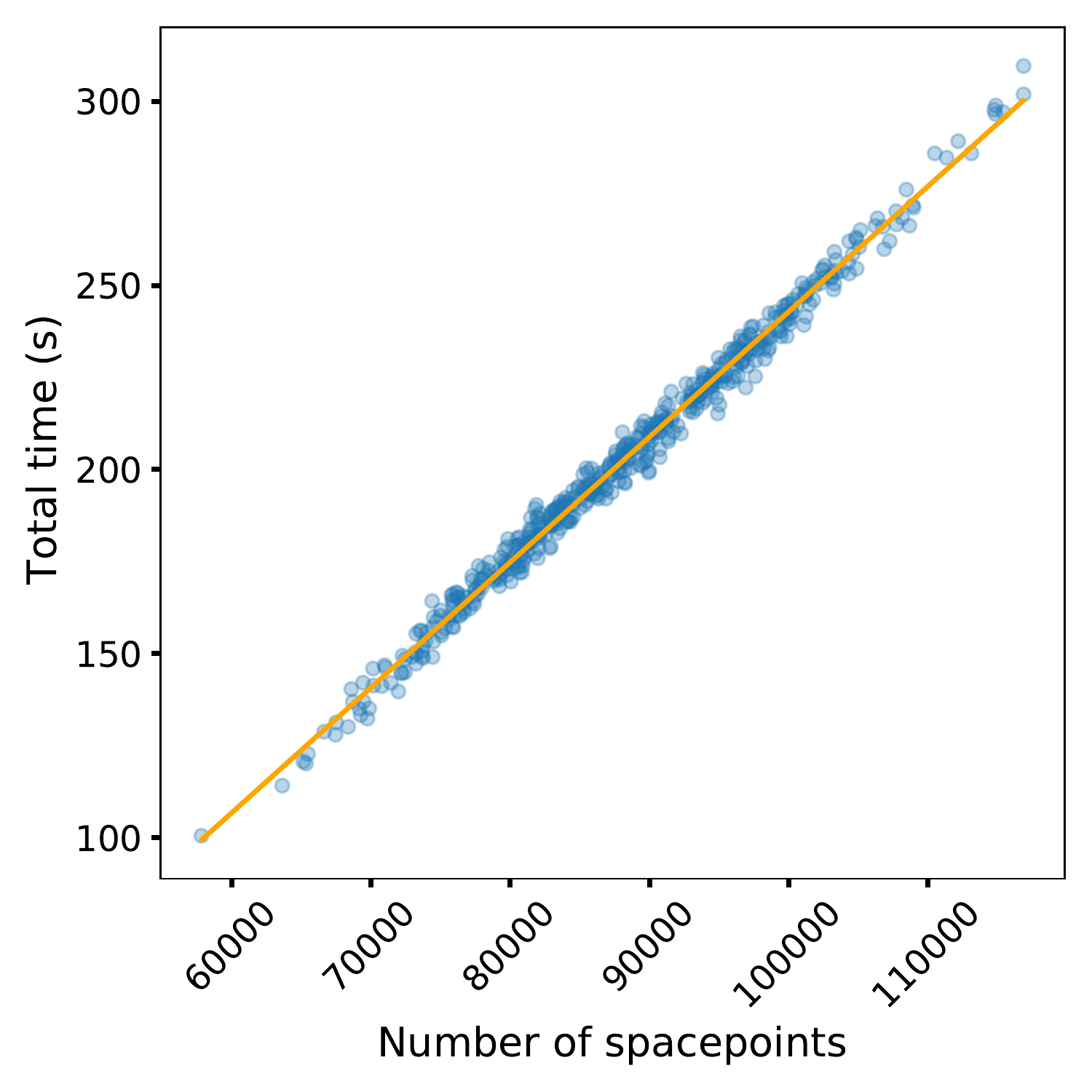}
    \includegraphics[width=0.9\textwidth]{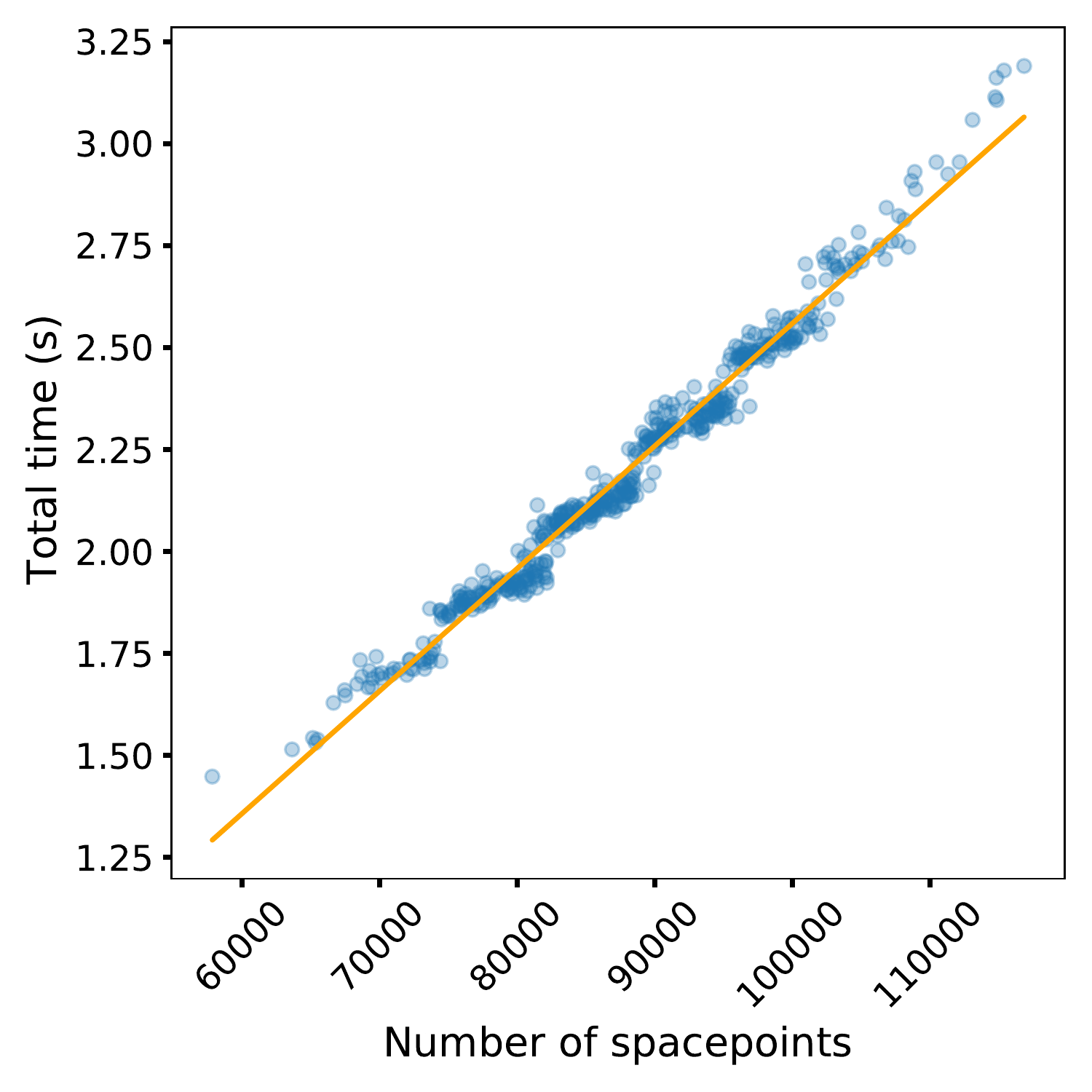}
    \caption{Total inference time as a function of number of spacepoints in each event for CPUs (top) and GPUs (bottom).}
    \label{fig:inf_timing_vs_nhits}
\end{figure}

In addition, Figure~\ref{fig:inf_timing_vs_nhits} shows how the total inference time depends almost linearly on the number of spacepoints in the event for both CPUs and GPUs. The step-like dispersion in the GPU case is due to the splitting of the inputs to the filtering step into mini-baches. A step-like jump indicates one more mini-batch is added.

Many optimizations were introduced to the pipeline in order to achieve these GPU timings, which before optimization took over 20 seconds per event. These improvements include porting all data processing to the GPU-accelerated CuPy library~\cite{Okuta2017CuPyA}, writing custom sparse operations for graph processing (e.g. doublet-to-triplet conversion~\cite{Fey/Lenssen/2019}, graph intersection methods), using FAISS~\cite{JDH17} for large-k NN graph construction, and performing track labelling with CuGraph's connected component algorithm on GPU~\cite{cugraph}\footnote{on CPU, track labeling uses the DBSCAN algorithm~\cite{DBSCAN}.}. These improvements are specific to the inference stage; training optimizations will be discussed in the following section, and ongoing developments in \S~\ref{sec:future_work}. No CPU-specific optimization was performed in this work.

\section{Discussion}
\label{sec:discussion}

The performance given above is the result of experimentation across various feature sets, architectures, model configurations and hyperparameters. It has also been necessary to overcome a variety of training hurdles in terms of memory and computational availability. We describe here training and inference details that should allow a reader to reproduce these results on the provided codebase. 

\subsection{Feature Set}

The input dataset includes both spatial coordinates and highly granular pixel cluster shape information. Graph construction (the second pipeline step in Figure~\ref{fig:pipeline}, that includes learned embedded space model and edge filter model) appears to benefit significantly from the cluster shape information, approximately doubling the purity for a held fixed high efficiency. The summary cluster shape statistics include the number of channels and the total charge deposited, as well as local and global representations of the cluster as a high-level feature vector. Details about the calculation of this feature vector as well as a
thorough exploration of the effect of cluster shape information on seeding performance are provided in Ref.~\cite{fox20204d}. Cluster shape information does not appear to improve the performance of the GNN, and in fact seems to degrade it. This suggests that the width of the GNN hidden layers is not great enough to capture the functional relationship of cluster information between nodes. Scaling to a width that properly explores this question would require more memory than available on the Nvidia A100 GPUs used for this study.

Depending on the final goal of the pipeline, further features can be included in the loss calculation in order to bias the model towards desired regions. For example, if our aim is to maximize the TrackML score (described in Ref.~\cite{TrackMLAccuracy2019}) –- a weighting function $s_i$ that places more importance on a spacepoint $i$ from a longer and higher \pt\ track, and in the first and last sets of detector layers -- we can weight-up true edges by this function, normalized to have a mean of weight $=1$. To measure the performance of models trained to this goal, we introduce a \textit{weighted} purity measure. Weighted purity is defined as a function the TrackML weights $w_{ij}$ and the truth $y_{ij}\in \{0,1\}$ of each edge connecting spacepoint $i$ and spacepoint $j$,

\begin{multline}
    \textnormal{Purity}_\textnormal{weighted} = \frac{\sum_{ij} w_{ji} y_{ij}}{\sum_{ij} w_{ij}},\\  w_{ij} = \begin{cases} \frac{1}{2}(s_i + s_j), \textnormal{ if } y_{ij} = 1 \\ 1, \textnormal{ if }  y_{ij} = 0
    \end{cases}
\end{multline}

We see significant improvements in this metric when validating on the weighted model: the Embedding Network improves from a weighted purity of $1.7\% \pm 0.2\%$ to $2.0\% \pm 0.3\%$, while the Filter Network improves from a weighted purity of $8.4\% \pm 0.6\%$ to $11.7\% \pm 1.0\%$. Given this weighting, the model learns to prioritize higher \pt~ and longer tracks, while disregarding less informative tracks. Using this bias, we can achieve the same TrackML score with a constructed graph size reduced by approximately 25\%. Using this technique to improve the TrackML score is an ongoing work.

\subsection{Graph Construction}

Having chosen a feature set, to train the learned embedding space we use a training paradigm commonly referred to as a Siamese Network~\cite{Chicco2021}, where a particular spacepoint - called the \textit{source} - is run through an MLP, here 6 layers each with 512 hidden channels, hyperbolic tan activations, and layer normalization. The final layer of the MLP takes the features to an 8-dimensional latent space. A different, comparison spacepoint - called the \textit{target} - is also run through this same Embedding Network, and the L2 norm distance $d$ in the latent space between the source and target enters a comparative hinge loss
\begin{align}
    \mathcal{L}_\textnormal{hinge} = \begin{cases} 
    d^p, \textnormal{ if } y_{ij} = 1 \\
    \textnormal{max}(0, 1 - d^p), \textnormal{ if } y_{ij} = 0
    \end{cases}
\end{align}
 where $p$ is a hyperparameter that we choose to be $2$. 
 
If the source $i$ and target $j$ spacepoints share an edge in the event’s truth graph~\footnote{one can also designate $y_{ij}=1$ for source and target in the same track, rather than \textit{immediate neighbors} in the track. This does lead to similar performance in later stages of the pipeline, but the more lax concept of truth leads to graphs around three times more dense than the strict track neighbor definition.}, we designate them as neighbours with $y_{ij} = 1$, otherwise they are designated $y_{ij}=0$. In this way, the hinge loss draws together truth graph neighbors and repels non-neighbors.  

Training performance of the Embedding Network is highly dependent on choice of source-target example pairs. In early epochs, it is enough to choose random pairs. However, at some point, many random pairs will contribute no gradient to the loss, as they will be separated by a distance greater than the margin. At that point, it is useful to implement hard negative mining~\cite{hnm}. We run a GPU-optimised k-nearest-neighbor (KNN) algorithm~\footnote{We use two high-performance libraries, FAISS~\cite{JDH17} and Pytorch3D~\cite{ravi2020pytorch3d}, depending on number of nearest neighbors $k$. Fastest performance is obtained with FAISS for $k > 35$, Pytorch3D for $k <=35$.} to mine examples around each source vector, within the hinge margin $d=1$. The computational overhead of the KNN step is significantly offset by the examples mined which all contribute to the loss. 

A similar technique is used in the Filter Network, where the vast majority of the edges produced from the graph construction in the embedded space are easy to classify as fake. This is already a highly imbalanced dataset, with around 98.5\% of edges fake. Again, within several epochs, the Filter Network is able to classify many of these as fake, so we balance each batch with all true edges, the same number of hard negatives (i.e. negatives the filter is unsure of) and the same number of easy negatives (to maintain performance on these edges). The Filter Network is a  MLP that takes the 24-feature concatenated edge features and feeds forward through 3 layers of 1024 hidden channels, to a binary cross-entropy loss function. 

\subsection{GNN Edge Classification}

In choosing the best GNN architecture, memory usage remains a significant constraint. The Interaction Network (IN)~\cite{interaction-networks} presented in these results does appear to marginally attain the best performance against Attention Graph Neural Networks (AGNN)~\cite{velivckovic2017graph,heptrkx-ctd2018} – the other class of GNN considered for the pipeline. However, both of these networks require gradients to be retained in memory for every graph edge. Indeed, this anisotropic treatment of edges (i.e. a node is able to receive the messages of each of its neighbors in a non-uniform way) is what allows these two architectures to be so expressive. Depending on hardware availability, we have found two solutions to the memory constraint. Access to next-generation Nvidia A100 GPUs allowed an IN to be trained with 8 steps of message passing, aggregating edge features at each node, and each node and concatenated edge features passing through two-layer MLPs of [128, 64] hidden features and ReLU activations~\cite{pmlr-v15-glorot11a}. Choice of aggregation function should be permutation invariant. In this work, we take it to be a summation.

For lower-memory GPUs, such as the Nvidia V100, we attained similar performance training the AGNN architecture, with [64, 64, 64]-channel MLPs applied to each edge and node. Adding residuals~\cite{ResNet} across the 8 message passing steps greatly improved performance in this case. To fit full-event training on a single V100, it was necessary to employ various techniques, such as mixed precision training and gradient checkpointing. The latter stores only the input of each layer, not the gradients. On the backward pass, gradients are re-calculated on the fly, allowing for a ~4x reduction in memory usage for an 8-iteration GNN. Another technique explored is to split the events piecemeal and train on each piece as a standalone batch. There is a noticeable impact on performance due to messages being interrupted at the graph edges. In future work, we will present ongoing efforts to parallelise these graph pieces across multiple GPUs, retaining the high performance that full-event training allows. 
 
\subsection{Physics-inspired data augmentation}

Preliminary work on using coordinate transforms to augment the training data has been explored with varying degrees of success. In this study, focused on track seeding, only the innermost detector layers (volumes 7-9 in Figure~\ref{fig:trackml_detector}) were used. 

One promising approach is to make a copy of each graph in the training set that has been reflected across the phi-axis~\cite{DBLP:journals/corr/abs-1712-04621}. 
The phi reflection creates the charge conjugate graph and helps to balance any asymmetry between positive and negatively charged particles within the training set. Using the phi-reflected graphs boosts efficiency by $\simeq 2\%$ and purity by $\simeq 1\%$ in the barrel. This performance boost comes at the cost of doubling the training time.  In future work, we will investigate the opportunity of integrating charge conjugation symmetry into the network itself.

A second promising trick is to use a Hough Transform~\cite{hough_transform,Gradin:2017jxr} on the graph to create edge features. Using the Hough parameters as edge features boosts efficiency by $\simeq 2\%$ and purity by $\simeq 1\%$. A further efficiency boost of $\simeq 3\%$  (and $\simeq 2\%$ to purity) comes from using the Hough accumulator to extract an edge weight. This edge weight effectively pools information from every node, and therefore comes at a large computational cost  (filling the accumulator in Hough space). On the other hand, the Hough parameters can be computed quickly from the two nodes that define the edge. 

\section{Conclusions and Future Work}\label{sec:future_work}
This works shows how a tracking pipeline based on geometric deep learning can achieve state-of-the-art computing performance that scales linearly with the number of spacepoints, showing great promise for the next generation of HEP experiments. The inference pipeline has been optimized on GPU systems, on the assumption that the next generation of HEP experiments will have widespread access to accelerators either locally in heterogeneous systems  \cite{Pata:2021oez,fahim2021hls4ml} or remotely \cite{Krupa_2021,kuznetsov2020mlaas4hep}. 

Within the simplifying assumptions of the TrackML dataset, we have shown how the Exa.TrkX pipeline could meet the tracking performance requirements of current collider experiments. Preliminary studies suggest that this performance should be robust against systematic effects like detector noise, misalignment, and pile-up.

Much remains to be done to validate these promising results. To this end, the Exa.TrkX project is collaborating with physicists from ATLAS~\cite{atlas}, CMS~\cite{cms}, DUNE~\cite{DUNE2018}, ICARUS~\cite{Bagby_2021}, and MuonE~\cite{Abbiendi2017}.

The goal is to adapt the Exa.TrkX pipeline to each experiment's needs and simulated datasets, measure its performance and robustness against systematic effects according to the experiment metrics. For example, it is crucial for HL-LHC experiments to study the performance of tracking algorithms in dense environments, like high-\pt\ jets. Given the interest in long-lived particle observation at the HL-LHC, it will also be important to study the performance of the Exa.TrkX pipeline for tracks coming from a displaced vertex~\footnote{it may be worth noticing that in LArTPC applications~\cite{hewes2021graph} all tracks come from a displaced vertex.}.

On the computational side, there are several optimization opportunities to explore  systematically, including mixed precision training, multi-GPU training and inference with graph data parallelisation (that is, one event spread across multiple GPUs)~\cite{Scardapane_2021}; locality sensitive hashing to speed-up KNN/graph construction stage~\cite{10.1145/276698.276876}, model quantization, operator fusion and other improvements with TensorRT~\cite{tensorrt}, clustering of final node embeddings rather than hard connected components method with GravNet-style architectures~\cite{Qasim:2019otl}.

The distributed training results presented in this work are promising but still preliminary. To fully exploit the capabilities of upcoming HPC systems and to further reduce training time while potentially pushing further on model size, it will be beneficial to perform further studies on large scale training of GNNs for track reconstruction. Given the size of the input graphs, this problem may be amenable to training techniques which parallelise the processing of input graphs across multiple GPUs in training.

Finally, it will be interesting to measure the computing performance of (parts of) the Exa.TrkX pipeline on domain-specific accelerators like Google TPU~\cite{tpu-eval} and GraphCore IPU~\cite{jia2019dissecting}, comparing power consumption, latency and throughput with "traditional" GPUs.

\section{Software availability} 
\label{sec:code}

A growing number of groups are studying the application of graph networks to HEP reconstruction (see \cite{duarte2020graph} for a recent review). Some of these works~\cite{biscarat2021realistic,Pata:2021oez,hewes2021graph,Heintz:2020soy,fox20204d,amrouche2021hashing,Tuysuz:2020eaa} have strong connections with the Exa.TrkX project. To promote collaboration and reproducibility, the Exa.TrkX software is available from the HEP Software Foundation’s Trigger and Reconstruction GitHub\footnote{\url{https://hsf-reco-and-software-triggers.github.io/Tracking-ML-Exa.TrkX}}. A pipeline of re-usable modules is implemented within the Pytorch Lightning system, which allows for uncluttered and simple model definitions. As each stage of the pipeline is dependent, logging utilities are integrated that allow a specific combination of stages and hyperparameters to be trackable and reproducible. Extensive documentation is provided to help track reconstruction groups start exploring geometric learning. The roadmap for this repository includes adding performance metrics to the codebase; a taxonomy of model features; and short tutorials in each of the available applications.

\section*{Acknowledgements}
This research was supported in part by: 
\begin{itemize}
    \item the U.S. Department of Energy’s Office of Science, Office of High Energy Physics, under Contracts No. DE-AC02-05CH11231 (CompHEP Exa.TrkX) and No. DE-AC02-07CH11359 (FNAL LDRD 2019.017); 
    \item the Exascale Computing Project (17-SC-20-SC), a joint project of DOE's Office of Science and the National Nuclear Security Administration;
    \item the National Science Foundation under Cooperative Agreement OAC-1836650. 
\end{itemize}

This research used resources of the National Energy Research Scientific Computing Center (NERSC), a U.S. Department of Energy Office of Science User Facility located at Lawrence Berkeley National Laboratory, operated under Contract No. DE-AC02-05CH11231. We are grateful to Google Co. for providing early access to Nvidia A100 instances in the context of the US ATLAS/Google Cloud Platform collaboration.

Finally, we thank Marcin Wolter (IFJ PAN), Ben Nachman, Alex Sim and Kesheng Wu (LBNL) for the useful discussions.



\bibliography{main,trackml}
\bibliographystyle{utphys} 

\end{document}